\documentstyle[psfig]{mn}
\title[Extremely Red Galaxies in the CDFS/GOODS field]
{The Nature, Evolution, Clustering and X-ray Properties of
 Extremely Red Galaxies in the CDFS/GOODS field }  
\author[N.D. Roche, O. Almaini, J.S. Dunlop]
{Nathan D. Roche$^{1,2}$, James Dunlop$^{1,3}$, and Omar Almaini$^{1,4}$\\
$^1$Institute for Astronomy,
     University of Edinburgh,
     Royal Observatory, 
     Edinburgh EH9 3HJ,
     Scotland.\\
{$^2$ \verb"ndr@roe.ac.uk"}\hspace{8mm}   
{$^3$ \verb"jsd@roe.ac.uk"}\hspace{8mm}
{$^4$ \verb"omar@roe.ac.uk"}\hspace{8mm}
}

\begin{document}

\maketitle
\begin{abstract}
We identify a very deep sample of 198  extremely red objects (EROs) in the 
Chandra Deep Field South, selected on the basis of
 $I_{775}-K_s>3.92$, to a limit $K_s\simeq 22$
using the public ESO/GOODS survey.

The ERO number counts flatten from a slope of $\gamma\simeq 0.59$ to 
 $\gamma\simeq 0.16$ at $K>19.5$, where they remain
below the predictions for pure luminosity evolution, and fall below
even a non-evolving model. This suggests there is a significant decrease with
 redshift in  the comoving number density of
passive/very red galaxies.

   We investigate the angular correlation function, $\omega(\theta)$,
of these EROs and detect positive clustering for
 $K_s=20.5$--22.0 sources. The EROs show stronger clustering 
than other galaxies at the
same magnitudes. The $\omega(\theta)$ amplitudes  are
 best-fitted by models in which the EROs have a comoving
correlation radius $r_0\simeq 12.5\pm 1.2~h^{-1}$ Mpc, or
$r_0\simeq 21.4\pm2.0~h^{-1}$ Mpc in a stable clustering model. 

We find a 40-arcsec diameter
overdensity of 10 EROs, centered on the {\it Chandra} X-ray source (and ERO)
XID:58. On the basis of colours we estimate that about 7, including
XID:58, belong to a cluster of EROs at $z\simeq 1.5$.

The 942 ksec {\it Chandra} survey detected 73 X-ray sources in the area
of our ERO sample, 17 of which 
coincide with EROs. Of these sources, 13 have 
X-ray properties indicative of
 obscured AGN, while the faintest 4 may be starbursts.
In addition, we find evidence that {\it Chandra} sources and EROs are
positively cross-correlated at non-zero
($\sim 2$--20 arcsec) separations, implying that they  tend to trace the same 
large-scale structures. 

In conclusion these findings  appear consistent with a
scenario  where EROs are the $z>1$ progenitors of  elliptical/S0
galaxies, some forming very early as massive spheroids, which are
strongly clustered and may evolve via an AGN phase, others more
recently from mergers of disk galaxies.

\end{abstract}

\section{Introduction}
The `extremely red objects' (EROs) are a population of very red
galaxies (e.g. with $R-K>5$ or $I-K>4$) which appear
 faintward of $K\sim 17.5$. EROs
are of special interest in the study of galaxy evolution, in that 
their colours and other properties suggest that they are 
the high-redshift  ($z\simeq 1$--2)
 counterparts and progenitors of local elliptical and S0 
galaxies, and are amongst
the oldest galaxies present at these redshifts.

Recent spectroscopy of some of the brightest EROs (Cimatti et al. 2002a;
Saracco et al. 2003) has
revealed a mixture of  old, passive galaxies (hereafter pEROs) and younger
star-forming or post-starburst galaxies with strong dusty-reddening
(hereafter dsfEROs). 
Some EROs also host AGN activity.  Recent {\it Chandra} surveys find that $\sim
10$--30 per cent of EROs on the
Chandra Deep Field North (CDFN) are detected in X-rays and, while in some
of these the X-rays may be produced by starbursts, the majority
have X-ray properties indicative of  obscured AGN (Alexander et al. 2002;
Vignali et al. 2003). 

In our previous paper (Roche et al. 2002, hereafter also Paper I), 
we combined our deep K-band survey of the `ELAIS:N2' field 
with the UFTI (reaching $K\simeq 21$) and Ingrid ($K\simeq 20$)
 cameras with a deep $R$-band image, to select a sample of
158 EROs as  $R-K>5.0$ galaxies.  The number counts of these EROs
 were lower than predicted by
a model in which all E/S0 galaxies evolve by `pure luminosity
evolution' (PLE), but could be fitted by a  model combining passive luminosity
evolution with galaxy merging and a decrease with redshift in the 
comoving number density of passive galaxies (`merging
and negative density evolution').
 
A number of studies have found that 
EROs are  more strongly clustered than other galaxies of 
similar magnitudes (Daddi et al. 2000; Firth et al. 2002).
As a measure of clustering, we
 investigated the angular correlation function, $\omega(\theta)$,  
of our ELAIS:N2 sample. Our results, combined with those of
Daddi et al. (2000) and Firth et al. (2002), were
best-fitted by models in which the EROs have 
a comoving correlation radius $r_0\simeq 10$--13 $\rm
h^{-1}$ Mpc - which  implies even stronger  intrinsic clustering than that of
present-day $L>L^*$ ellipticals ($r_0\simeq 8 \rm
h^{-1}$ Mpc).

 We examined the morphology of the brighter
($K\leq 19.5$) EROs (31) 
and estimated that $\sim 60$ per cent  were bulge-type 
(E/S0) galaxies with the others a mixture of 
disk, irregular and merging galaxies. This mix 
is consistent with the WFPC2 study of 
Moriondo, Cimatti and Daddi (2000). The angular sizes of the
EROs were generally consistent with our best-fitting evolution model, whereas
a non-evolving model underpredicted their 
surface brightness.

Seven of these 31 brighter EROs were detected as radio  sources
on a deep VLA survey to $F(1.4~{\rm  GHz})\simeq  27.6\rm \mu
Jy$. Six were point-like within resolution  1.4 arcsec.
 One of these was a powerful radio galaxy, and the other five were
much fainter radio sources within elliptical galaxies, 
 which could have been either obscured central starbursts or
 obscured AGN.

The seventh was an elongated radio source,  
aligned with an extended, low surface brightness
 optical morphology, suggestive of a dusty
starburst galaxy. This implied that some (but perhaps only a few per
cent) of EROs have 
star-formation rates as high as the
$>100$--1000 $\rm M_{\odot} yr^{-1}$ needed for radio 
detection in this survey. 
Smail et al. (2002) detected ${21\over 68}$ EROs on an even deeper VLA
image, and 
 found several other examples of dusty starburst galaxies with
extended and aligned radio emission.
  
A scenario was proposed in which the (i) pEROs are strongly
clustered primordial ellipticals which had formed as powerful sub-mm 
galaxies at even higher redshifts, (ii) the younger
 dsfEROs are starburst and
recent post-starburst galaxies formed from the merging of 
disk galaxies, which would only later become
 passive E/S0s. At all epochs, interactions cause  
new dsfEROs to be added to the
ERO class,  increasing the 
comoving number density (as in our M-DE model).

 If  the progenitors of dsfEROs
are less clustered  than the old pEROs, this process would be
accompanied by a dilution of ERO
clustering with time (i.e. an increase with redshift).
Daddi et al. (2002) have recently claimed that 
 spectroscopically
selected pEROs are indeed more clustered than dsfEROs. However, their 
sample was small (33) and included only the brightest EROs (mostly
$K<19.2$), and the clustering was analysed in redshift space only, 
so this needs to be verified with larger and deeper samples.

Another result of interest is the 
Almaini et al. (2003) finding that
sub-mm (detected with SCUBA) and X-ray ({\it Chandra}) sources on the
ELAIS:N2 field, were
positively cross-correlated at non-zero separations. This was interpreted
as possible evidence for an evolutionary connection between the two. 

In this scenario (Archibald et al. 2002; Granato et al. 2003), the
sub-mm sources are giant ellipticals in the process of formation. At
early epochs, the most massive galaxies would form stars very rapidly,
producing far-infra-red flux  (redshifted to the sub-mm). In  
parallel with this, a supermassive, dust-enshrouded black hole is
formed in the centre of the galaxy.
 After $\sim 1$ Gyr, the emission from the accreting
black hole becomes sufficient to
eject the dust and remaining
gas from the galaxy, halting further star formation,
 and the galaxy ceases to be sub-mm bright but instead is
transformed into an  X-ray luminous QSO.

The relevance of this to our present study of EROs is that, at
 the end of the QSO phase, the massive, gas-stripped galaxies can only  become 
passively evolving red ellipticals, i.e.,
EROs. One consequence of this  might be a
similar cross-correlation between EROs and unreddened, non-ERO 
{\it Chandra} sources.

The recent public release of the first installment of the ESO/GOODS
survey -- with optical and near-infra-red imaging of the Chandra Deep
Field South (CDFS) -- provided us with an excellent opportunity to
investigate both
the counts and clustering of EROs (to a fainter limit than previously) and
their association with a very deep sample
 of X-ray sources.
 
The layout of this paper is as follows: Section  2  
describes  the observational data and any data reduction, 
Section 3  the identification and colour-classification of the EROs, 
Section 4 
the ERO number counts, comparing these with evolutionary models, Section 5
the ERO angular correlation function, $\omega(\theta)$, with an 
 interpretation in terms of intrinsic clustering. In Section 6 we
investigate the numbers of close pairs of EROs and identify any larger
clusters, and in Section 7 identify EROs which are also {\it Chandra}
X-ray
sources and look for a correlation between these two classes.
Section 8 concludes the paper with a discussion of our findings. 

In Paper I we had assumed a cosmology with
$\Omega_M=0.3$,
$\Omega_{\Lambda}=0.7$ (modelled using the analytic form of Pen 1999)
and $H_0=55$ km $\rm s^{-1} Mpc^{-1}$. In view of new results from the 
Wilkinson Microwave Anisotropy Probe (e.g. Spergel et al. 2003), we
here retain this $\Omega_M/\Omega_{\Lambda}$,
 but adopt the higher $H_0=70$ km $\rm s^{-1}
Mpc^{-1}$, which 
reduces the time since the Big Bang from 17.16 to 13.48 Gyr.
 Some $H_0$-dependent quantities are given in units of 
$h=(H_0)/100$ km $\rm s^{-1} Mpc^{-1}$ )

\section{Observational Data}
\subsection{Near Infra-red ($JHK_s$) Imaging}
The Great Observatories Origins Deep Survey (GOODS) (Dickinson and
Giavalisco 2003) is a public,
multiwavelength survey of the Hubble Deep Field North and the CDFS.
 The ESO contribution to GOODS 
will include deep near-IR and optical imaging of the whole 150
$\rm arcmin^2$ of the CDFS. The near-IR observations are
being carried out using the  `ANTU' VLT (Very Large Telescope), 
with the Infrared Spectrometer and Array Camera
(ISAAC). ANTU was
the first of four 8m
telescopes installed at the European Southern Observatory (ESO)
 at Cerro Paranal, Chil\'{e}. ISAAC is equipped with a Rowell
 Hawaii $1024\times 1024$ pixel Hg:Cd:Te array
covering 1--$2.5\mu m$, with a pixelsize 0.1484 arcsec giving a $2.5\times 2.5$
arcminute field-of-view. 

ISAAC observations will eventually cover the whole CDFS,
 as a 32-frame mosaic
in the $J$ ($\lambda_{central}=1.25\mu m$) $H$ ($1.65\mu m$) and $K_s$
 ($2.16\mu m$) passbands. 
ISAAC observations of the first 8 GOODS
fields, covering a contiguous area near the centre of the CDFS
(approx. RA $3^h 32^m 30^s$, Dec. $-27:47:30$) 
were completed on 1 February 2002 and made publicly
available on 9 April 2002. 
\subsection{$I_{775}$-band Imaging with the HST ACS}
Another component of the GOODS survey is deep optical 
imaging of the CDFS with the Advanced Camera for Surveys (ACS) on the
Hubble Space Telescope (HST). The ACS produces a $4096\times 4096$
pixel image with pixel size 0.05 arcsec. We make use here  of newly
released data from four
phases of ACS observation, obtained from 3 October to 22 December
2002, which cover the CDFS in mosaics of either 15 or 16 fields.
The CDFS imaging was performed in four
 ACS passbands ($BVIZ$). In this paper we make
use of only the $I_{775}$-band imaging
($\lambda_{pivot}=0.769 \mu m$), where the exposure time per field
is 1040 seconds for each phase.

\subsection{{\it Chandra} X-ray Observations}
We also make use of the CDFS X-ray source catalog of Giacconi et
al. (2002). The CDFS was observed with the {\it Chandra} X-ray
Observatory for eleven pointings during 1999--2000, which were combined to
give 942 ksec exposure time. The {\it Chandra} aim-point of
RA $3^h 32^m 28^s$, Dec. $-27:48:30$, where sensitivity and resolution
are optimal, is approximately concentric with
our ISAAC data. 

Giacconi et al. (2002)
detected a total of 346 sources in a soft (0.5--2.0~keV)
and/or in a hard (2--10~keV) X-ray band, 
reaching aim-point  detection limits
$F(0.5$--$2.0~{\rm keV})=5.5\times 10^{-17}$ and 
$F(2$--$10~{\rm keV})=4.5\times 10^{-16}$ erg $\rm s^{-1} cm^{-2}$,
with  resolution $\rm FWHM\simeq 0.7$ arcsec.
Of these sources, 73 are within the area of our 
ISAAC data. 
\section{Reduction and Analysis}
\subsection{ISAAC $JHK_s$ Data}
The 8 fields of ISAAC data were obtained from ESO (www.eso.org) in a reduced
form, as 24 images (8 fields $\times$ 3 passbands)
  each with a weighting map, astrometry and
 photometric zero-point. Total exposure times in each field 
were 20.8--28.6 ksec in
$K_s$, 13.92--18.12 ksec in $H$  and 7.92--15.12 ksec in $J$. 

For our
purposes the only further processing required was to mosaic the 8 fields into
a single image for each of the three passbands.
 However, there was considerable variation between these in
signal-to-noise and in zero-point, and to take this into account:

(i) we scaled the images and their weighting maps to a common photometric 
zero-point, that of the first
field (the scaling of a weighting map is the inverse of that applied
to the data image). Zero-points were also corrected
 for the Galactic extinction on the CDFS,
but this is only 0.008 ($J$), 0.005 ($H$) and 0.003 ($K_s$) mag.

 (ii) We multiplied the images by their weighting
maps. The resulting products were then  mosaiced using {\sevensize
IRAF} `combine' with `offsets=wcs' (using the supplied astrometry
which is sufficiently accurate), and `scale=none'. Note that this
averages input images where they overlap in area. 

(iii)  The weighting maps 
were mosaiced in the same way.

 (iv) the mosaic from (ii) was then divided by the mosaic from
(iii). The result of this is a `final' mosaic where the contribution
at each pixel from each of the input images is weighted by its
appropriate weighting map. The division by the mosaiced weighting map
 normalizes the sum of these input image weightings
to unity at each pixel. Hence the zero-point of the final mosaic
should be uniform across all pixels and the same as
that of the input images. 

 (v) We still need  a weighting map for the final mosaic.
To obtain this the mosaiced
weighting map from (iii), representing the average depth of the input
images contributing at each pixel, is multiplied  by a map of 
the number of these contributing images, thus giving a map of 
the total depth
of observation at each pixel.

The final mosaics for each of the three passbands, with their
respective weighting maps, were then ready to be used for source
detection.
The mosaicing procedure was checked by measuring source magnitudes on 
individual unmosaiced frames and comparing with those
from the mosaic, and by checking for any sign of positional 
mismatch in overlap regions.

\subsection{ACS $I_{775}$-band Data}
The ACS data was obtained from the HST archive
 (archive.stsci.edu/hst/goods.html), already flat-fielded and
 calibrated to a common zero-point. The first and third (and the
 subsequently released fifth) observation phase covered the CDFS in a
 $3\times 5$ mosaic of 15 pointings, with the long axis on position angle 
$-22.35{\degr}$. However, the second and third phase covered the CDFS
 in a 16 field grid  orientated at $45{\degr}$ to this, ie. with
 position angle $-67.35{\degr}$. 

The number of pixels was too great for the whole observed area to be
mosaiced into a single image. Instead, the four phases of observations were
 combined into 15 frames,
corresponding to the 15 field areas of the first phase of
observation. Because of the two different orientations of the data, for each
of these 15 frames there were typically 8 observed images which
covered all 
or part of the field area. These data were combined by much the same
methods as the ISAAC data above, but with an additional step of
rebinning:

(i) For each of the 15 frames, {\sevensize IRAF} `wcsmap' was used to fit
 transforms between the pixel grid of the
phase one observation and that of all subsequent observations
 (typically 7) which
 cover at least part of the same area of sky. {\sevensize IRAF} `geotran'
 could then be used to rebin these observations into the same pixel
 grid as the phase one observation (which was not rebinned). This
 rebinning would allow observations taken at different position angles to be
 simply added together. 

(ii) The exposure maps supplied with each observed image were
similarly rebinned into the phase 1 pixel grid.

(iii) For each of the 15 frames, 
the data images to be combined (rebinned, except for
phase 1) were multiplied by their respective
 exposure maps and then added together using {\sevensize
IRAF} `combine'. The exposure maps were similarly combined. Then 
the combined data$\times$exposure map image was divided by the
 combined exposure map,  to produce
a `final' co-added image 
  where the contribution at each pixel 
from each of the input images is proportional to its weighting (from
its exposure map), and the sum of the weighting factors is normalized
to unity.

(iv) A weighting map for each of the 15 `final' frames was produced by
multiplying the combined weighting map by a map of the number of input images
contributing at each pixel (again, as for the ISAAC data).

\subsection{Source Detection}
Sources were detected on the ISAAC and ACS images using
`SExtractor'(Bertin and Arnauts 1996),
 as in Paper I. We make use of both `total' magnitudes,  derived by
 `SExtractor' by fitting elliptical apertures to each detection, and 
aperture magnitudes, measured by `SExtractor'
 in circular apertures of fixed 2.0 arcsec diameter. The former will be
used for limiting the sample and the latter for measuring  colours.
Magnitudes are given in the Vega system and can be converted to the AB
system as $(K_s,H,J,I_{775})_{AB}= (K_s,H,J,I_{775})_{Vega}+(1.841, 1.373,
0.904, 0.403)$.

Source detection was performed on the ISAAC $K_s$ mosaic (together
with its weighting map), with the chosen
criterion that a source must exceed $1.4\sigma_{sky}$ above the
background, or 22.75 $K_s$ mag $\rm arcsec^{-2}$,
 in at least 6 contiguous pixels. In addition,
a detection filter (3.0 pixel FWHM Gaussian) was employed. Detections on
the low signal-to-noise edges of the mosaic were excluded,
 leaving a data area of 50.4 $\rm arcmin^2$. 
   The catalog of $K_s$ detections forms the source list for our ERO sample.  

Photometry of these sources in $H$ and $J$ was obtained simply by positionally
registering (to the nearest pixel) the $H$ and $J$ mosaics to the
 $K_s$ mosaic and then running SExtractor in `double-image mode' to
detect sources in $K_s$, as before, and then measure their fluxes on the
$H$ and $J$ images.
Sources were detected separately on the ACS $I_{775}$-band image with the
criterion that they exceed $1.75\sigma_{sky}$ in at least 8 pixels, and 
a 2.5 pixel  FWHM Gaussian detection filter.
\subsection{$K_s$-band Galaxy Counts}
As some detections, especially at brighter magnitudes, will be
Galactic stars, we performed star-galaxy separation on the 
source list, using a plot of
peak/total $K_s$ flux  against magnitude. This was effective to a limit
 $K_s=19.0$, to which  
30 objects were classed as stars, and all fainter detections were
assumed to be galaxies. We estimate the resolution of the ISAAC data
as the mean Gaussian FWHM of the non-saturated stars, 0.46 arcsec.
At this depth, the
 slope of the Galactic star counts is only ${dN\over dm}\simeq 0.1$,
(e.g. Roche et al. 1999) 
which by extrapolation would indicate the star-contamination at
$19<K_s<22$ to be $\sim 63$ stars, or 4.6 per cent of the galaxy sample.

Figure 1 shows the differential number count of galaxies as a function
of total $K_s$ mag, shown with other counts and two models from Paper I.
Saracco et al. (2001) also observed in the $K_s$ band, and the $2.2\rm
\mu m$ results of Paper I and Moustakas et al. (1997) are plotted
assuming a small colour term $K_s-K\simeq 0.04$, 
estimated for a passive ERO model at $1<z<2$.
  
Our CDFS galaxy counts 
 agree well with the Paper I ELAIS:N2 counts 
to $K_s\simeq 21$, and reach $\sim 1.5$ mag deeper, turning over only at
$K_s>22$. Comparison with the two even deeper surveys suggests we
are near-complete to $K_s\simeq 21.5$ with some ($\sim
20$ per cent) incompleteness at $21.5<K_s\leq 22$. In most of the
 analyses of this paper we
 consider samples limited at total magnitudes of 
$K_s=22.0$ and $K_s=21.5$, for
 the more
 incompleteness-sensitive measures giving primary
 emphasis to the $K_s\leq 21.5$ sample. 
 \begin{figure}
\psfig{file=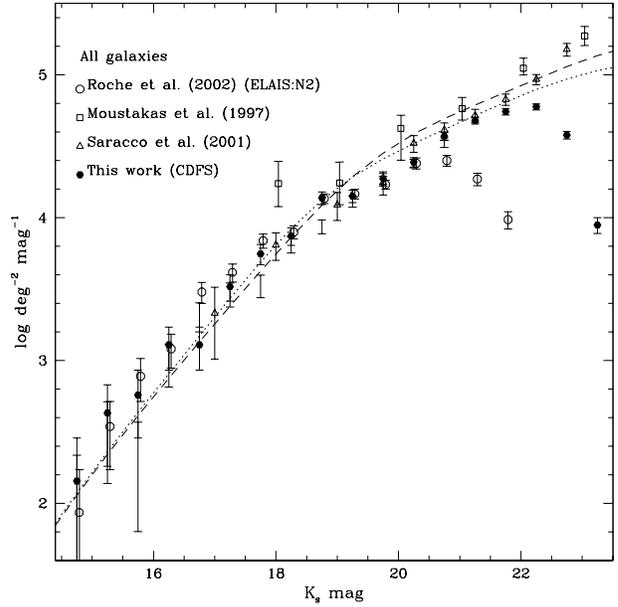,width=90mm}
\caption{Differential number counts of galaxies against total $K_s$
mag, for  the ISAAC CDFS and ELAIS:N2 (Paper I) fields
(shown well beyond completeness to illustrate the difference in
depth),
 and from the even deeper surveys of Moustakas et al. (1997) and
Saracco et al. (2001) ($K$-band observations are plotted assuming 
$K_s-K=0.04$).
Also plotted are  PLE (dotted) and
$R_{\phi}=R_{m^*}=0.3$ merging (dashed) models (as in Paper I but with
$h=0.7$.)}
\end{figure}

\section{Selection of the ERO sample}
\subsection{Identifying the EROs}
EROs are essentially galaxies with the red colours expected for  high-redshift
ellipticals. In  Paper I they were selected on the basis of a colour
$R-K>5.0$, while  some authors have used
slightly stricter criteria of $I-K>4.0$ or
$R-K>5.3$ (e.g. Mannucci et al. 2002).
 Throughout this paper we select EROs as $I_{775}-K_s>3.92$, which
corresponds to  our evolving 
E/S0 model at $z>0.99$ and is approximately equivalent (on the
 basis of a passive ERO model) to the $R-K>5.0$ of Paper I 
and to $I-K_s>3.75$
for the slightly longer $I$-band (eg. as on the VLT FORS1 instrument).

 The first step in identifying the EROs was to
match each object in the $K_s$ source list with its closest
counterpart on the ACS $I_{775}$ image,
 using the RA and Dec co-ordinates given by the
astrometry supplied with these images. We allowed offsets up to a  
maximum 1.0 arcsec radius.
Matching the $K_s$ source list to their $H$ and $J$
counterparts was more straightforward, as the use of double-image
mode meant there were no positional offsets.

ERO selection was initially performed on a subset of the data and
 the candidate EROs examined by eye to check on the reliability of
 the procedure. The selection method was then modified where necessary.

A list of EROs, numbering 155, was then selected 
as all galaxies detected in both $K_s$ and $I_{775}$,
 with total magnitude 
$K_s\leq 22.0$ and colour (from the 2.0 arcsec diameter aperture
magnitudes) $I_{775}-K_s>3.92$. Fixed-aperture colours
 were found to be more reliable for ERO selection
than colours from total magnitudes, the reason being that
 in a significant fraction of
galaxies SExtractor measured total
magnitudes within very different sized apertures for 
$K_s$ and $I_{775}$, giving rise to errors and selection biases.
 The mean offset between the $K_s$ and $I_{775}$
centroids was only 0.225 arcsec. 
With this positional accuracy, and a background density of 
about 0.05 $I_{775}$ detections 
$\rm arcsec^{-2}$,
 the probability of chance superposition producing a false match
(closer than the true $I_{775}$ counterpart) is 
only $\sim(\pi 0.225^2 \times 0.05)$, 0.75 per cent, for each galaxy.

In addition, 53 galaxies with $K_s<22.0$  had
 no detection within 1.0 arcsec  on the ACS
data. Many of these will also be EROs, but not all -- some, for example, are 
confused  with neighbouring brighter objects in the shorter passband,
 causing their $I_{775}$ detection
centroid to be shifted outside of the matching radius. To 
`filter' these 53 objects, we used {\sevensize IRAF} `qphot' to measure
their  aperture
(2.0 arcsec diameter) magnitudes on the ACS images 
at the co-ordinates
positions given by  their ISAAC detections,   
and then only
accepted the objects with
$(I_{775}-K_s)_{aperture}>3.92$, which numbered  43. Visual checking
indicated
that this
method appeared to exclude the confused objects.

 Our ERO sample then consisted of a total of $155+43=~198$
 galaxies with $K_s\leq 22.0$, of
which 179 were brighter than the estimated completeness limit of
$K_s=21.5$. As none of the brighter ($K_s<19.0$)
 objects classed as Galactic stars had $I_{775}-K_s$ any 
redder than 3.0, 
 star-contamination of the ERO sample will be 
 zero or very small ($\leq 2$). 
\subsection{Colours of EROs}
ERO colours correspond to passive galaxies at approximately $z>1$,
where the the observer-frame $K$-band  and the $R$ or $I$-band will bracket
the large break at $4000\rm \AA$ rest-frame. However, any
type of galaxy, including a young starburst, could be this red if sufficiently
dusty. 

Figure 2a shows models of $I-K_s$
against redshift, which start forming stars 12.55 Gyr ago, at $z=6$.
We model galaxy spectral energy distributions 
 using a version of the
`Pegase' package (Fioc and Rocca-Volmerange 1997; see Paper I). 
The elliptical galaxy model
(solid line)
forms all its stars in a 1 Gyr burst at $6>z>3.4$, and the S0 model
with a 2 Gyr burst continuing to $z=2.6$. Both have $E(B-V)=0.4$
mag dust extinction during the bursts and subsequent
 unreddened passive evolution, and show 
ERO colours at $z>1$.
This choice of formation redshift appears reasonable in view of the
discovery of 
star-forming galaxies in all intervals of
redshift out to the current
maximum of $z=5.78$ (Bunker et al. 2003).

Dusty starburst models,
represented as  non-evolving starbursts observed at age 50 Myr, and 
heavily  dust-reddened by $E(B-V)=0.8$ and 1.0 mag,  are also plotted,
as dotted and dot-dash lines.
 These produce 
similar $I-K$ colours to passive galaxies, especially at $z>1$,
and if very dusty they could have $I_{775}-K_s>3.92$  at
$z\sim 0.5$--1.0. 
Older star-forming galaxies, or recent
 post-starbursts, would have intermediate spectra and could
be EROs with more moderate dust-reddening.

Figure 2b shows $I_{775}-K_s$ for all  $K_s\leq
 22$ galaxies, with larger symbols for the EROs, which, 
 as in Paper I, appear at $K_s\sim 18$. Most $I_{775}\leq 25.5$ EROs
 are detected on the
 ACS data, most at $I_{775}> 26$ are not.

\begin{figure}
\psfig{file=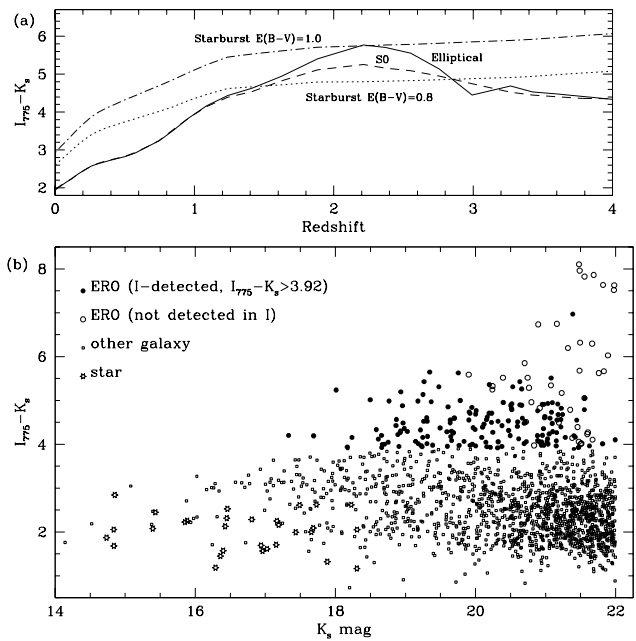,width=90mm}
\caption{(a) Observer-frame $I_{775}-K_s$ colour (Vega system) against
redshift for models representing evolving elliptical (solid) and S0 (dashed
lines) galaxies and dust-reddened young starbursts (dotted/dot-dash lines),
 (b) Observed $I_{775}-K_s$ (in 2.0 arcsec diameter apertures) for all $K_s\leq
22$ galaxies, with larger symbols for the 198 EROs.}
\end{figure}
\subsection{Colour classification}
It is possible, with the $H-K_s$ and $J-K_s$ colours 
provided by the ISAAC data, 
to achieve an approximate separation of passive and dusty,
 star-forming EROs (pEROs and dsfEROs). Pozzetti and
Mannucci (2000)
describe how these types can be separated on a plot of
$I-K$ against $J-K_s$, although this is only possible at
 $1\leq z\leq  2$, where the redshifted 
$4000\rm \AA$
break lies between the $I$ and $J$ bands. Here, we use essentially the 
same method, but also incorporate the $H-K_s$ colour, which extends
this method to $z\simeq 2.5$.

Figure 3 shows $I_{775}-K_s$ against $J+H-2K_s$ (ie. the sum of $J-K_s$ and
$H-K_s$) with the diagonal line showing the adopted pERO/dsfERO
divide, at $J+H-2K_s=0.86(I_{775}-K_s)-0.75$.

The majority of these EROs have colours consistent with either the passive
models at $1<z<2.5$ or the $E(B-V)\simeq 0.8$ starburst at $z<2$.
Using the plotted dividing line, we classify 109 EROs as pEROs and 89
as dsfEROs. It must be emphasised that these classifications
are subject to large random error, and many EROs lie close to the
dividing line, whether from scatter in
the magnitudes or to genuinely intermediate spectra.
To estimate the effectiveness of the classification we (using
the same method as Mannucci et al. 2002) calculate for each ERO,
the horizontal distance (i.e. $\Delta(J+H-2K_s)$ from the pERO/dsfERO 
dividing line on Figure 3, and divide this by the statistical error in
the EROs $(J+H-2K_s) $ colour.

 Many EROs are found to have
strong, $\sim 3$--$6\sigma$ classifications, but 47 classed as
pEROs 
and 55 dsfEROs are within $1\sigma$ of the line, and  hence about half of the
individual classifications (102) must be regarded as weak. The
resulting uncertainties 
in the total number of dsfEROs would be smaller by
$\sim \surd N$ and if we add this to the $\surd N$ of the total
number, we estimate there to be $89\pm \surd(55+89)=89\pm 12$ dsfEROs.
There would of course be an additional uncertainty from the
model-dependence of the dividing line. 
However, our estimated  dsfERO 
fraction,
$45\pm 6$ per cent, is  consistent with previous estimates of 
33--67 per cent
from direct spectroscopy  of $K_s<19.2$ EROs  (Cimatti et al. 2002a),
40--60 per cent   from multicolour photometry
of $K^{\prime}<20$ EROs (Mannucci et al. 2002), and 45--60 per cent
from combined radio observations and colours of  $K^{\prime}<20.5$ EROs
 (Smail et al. 2002).

\begin{figure}
\psfig{file=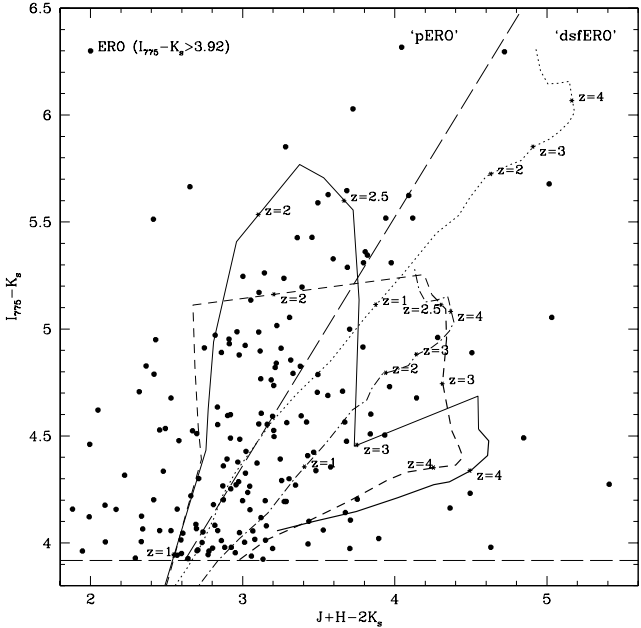,width=90mm}  
\caption{Plot of $I_{775}-K_s$ vs. $J+H-2K_s$ for all $K\leq 22.0$
EROs. Also plotted, with redshifts marked,  are the elliptical
(solid) and S0 (short-dashed) models, with 
heavier lines indicating the redshift range at which they are forming stars,
and starburst models with reddening of $E(B-V)=0.8$ (dotted) and 1.0
magnitudes (dot-dash). The long-dashed diagonal line shows the adopted 
separator of pERO and dsfERO types.} 
\end{figure}
\subsection{Distribution on the Sky}
\begin{figure}
\psfig{file=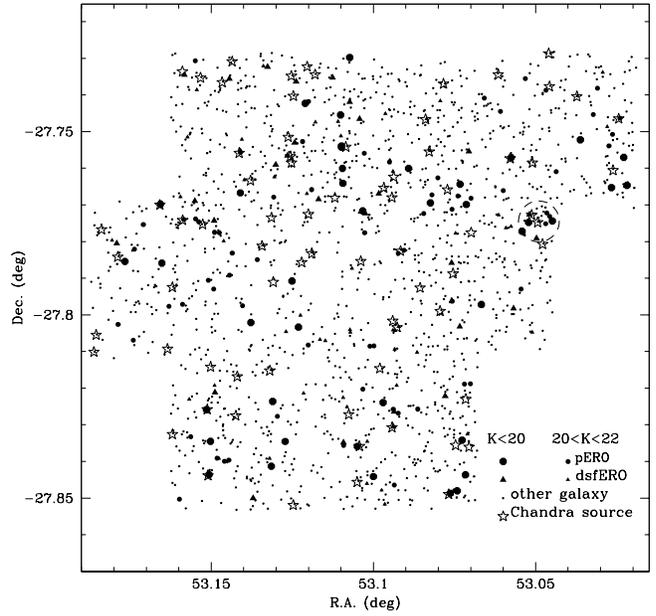,width=90mm}
\psfig{file=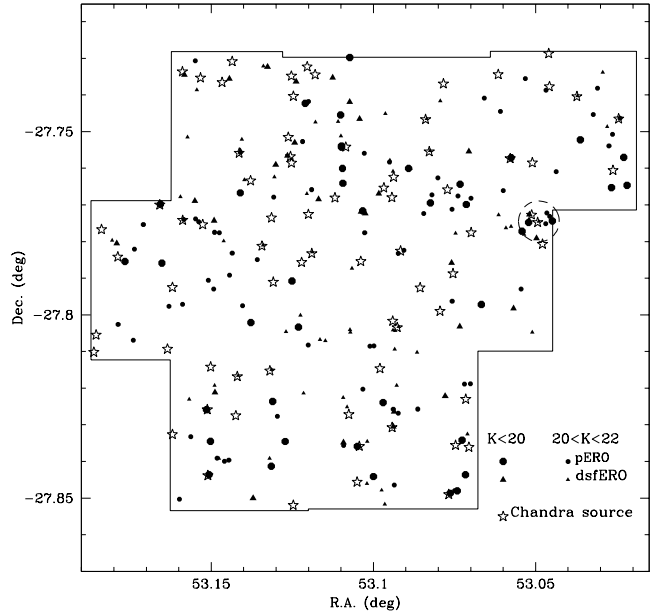,width=90mm}
\caption{(a) Distribution of colour-classifed pEROs and dsfEROs,
 other detected galaxies and {\it
Chandra} X-ray sources, on the area of the CDFS covered by the ISAAC
data. The dashed circle indicates a possible cluster of EROs centered 
on the {\it Chandra} source XID:58 (see Section 7.1).
(b) As (a) but with the field boundaries shown and the
non-ERO galaxies omitted, for clarity.}
\end{figure}
Figure 4 shows the distribution of the ERO sample, and the 73
 {\it Chandra} detections (Giacconi et al. 2002)  on the 
ISAAC mosaic area of the CDFS. 
The EROs appear somewhat clustered on small
scales, and a particularly 
obvious overdensity, centred on {\it Chandra}
source XID:58 (R.A. $3^h 32^m 11.85^s$ Dec -27:46:29.14), is
highlighted on the plot and discussed further in Section 7.1.

\section{Number counts of EROs}
\subsection{Observed counts on the CDFS}
The differential number counts for EROs on the CDFS are given in Table
1 and plotted on Figure 5. Firstly, the three ERO counts plotted here
are consistent within the error bars. Our CDFS counts are closer to
the uncorrected count of Paper I, suggesting the incompleteness
corrections may have been overestimated. For this paper,
 rather than attempting to
correct for incompleteness, we have assumed
completeness to $K_s=21.5$, and disregarded the final point 
in fitting slopes and models (incompleteness will appear more 
suddenly for this data due to its more uniform depth).

The most obvious feature on the plot is a
marked flattening of the count slope, seen in all three sets
of observations but most apparent with the CDFS. The count slope 
best-fitting the
the ELAIS:N2 and CDFS counts at $17.0\leq K_s\leq 19.5$ is
$\gamma=0.59\pm 0.11$, and  flattens to $\gamma=0.16\pm 0.05$ at $19.0\leq
K_s\leq 21.5$ (fit to CDFS counts only).

 \begin{table}
\caption{Observed number counts (number of galaxies
 $N_g$ and surface density
$\rho$ (in units $\rm deg^{-2}mag^{-1}$, with $\surd N$
errors) of EROs ($I_{775}-K_s>3.92$ galaxies) on the CDFS.}
\begin{tabular}{lcc}
\hline
\smallskip
$K_s$     & $N_g$  & $\rho$ \\
$<17.0$    & 0 & 0 \\
17.0--17.5 & 1 & $~143\pm 143$ \\ 
17.5--18.0 & 1 & $~143\pm 143$ \\
18.0--18.5 & 4 & $~572\pm 286$ \\ 
18.5--19.0 & 18 & $2573\pm 606$ \\
19.0--19.5 & 20 & $2859\pm 639$ \\
19.5--20.0 & 25 & $3574\pm 715$ \\
20.0--20.5 & 28 & $4003\pm 756$ \\
20.5--21.0 & 40 & $5718\pm 904$ \\
21.0--21.5 & 41 & $5861\pm 915$ \\
\hline
\end{tabular}
\end{table}
\begin{figure}
\psfig{file=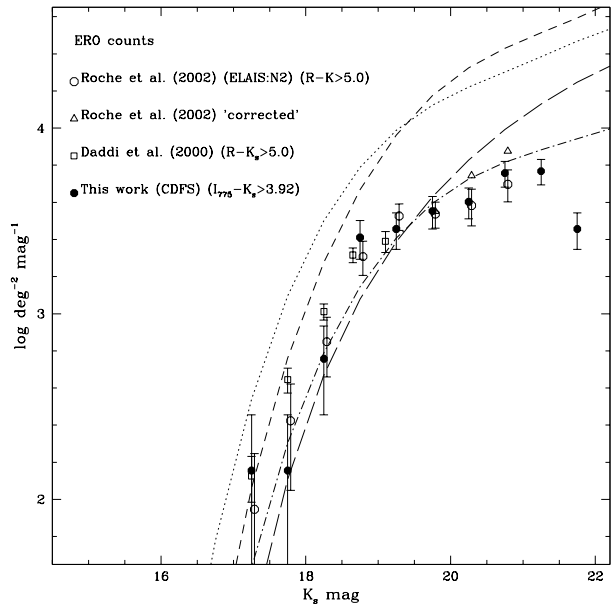,width=90mm}
\caption{$K$-band differential number counts for EROs selected as
$I_{775}-K_s>3.92$ on the CDFS and at
comparable limits of  $R-K>5.0$ on the ELAIS:N2 field
 (Paper I/Roche et al. 2002), and $R-K_s>5.0$ by  Daddi et
al. (2000). These are compared with the counts of  $I_{775}-K_s>3.92$ galaxies
predicted by PLE (dotted), non-evolving (long-dashed),
merging with $R_{\phi}=R_{m}=0.3$ (short-dashed), and `M-DE' merging with
$R_{m}=0.3$ $R_{\phi}=-0.49$ (dot-dashed) models. Error bars are $\surd N$.  }
\end{figure}

\subsection{ERO counts: Models and interpretation}
In Paper I we presented simple models of the number counts of EROs,
 based on the assumption that both pEROs and dsfEROs are
 progenitors of the present-day E/S0 galaxies, and 
hence that their luminosity function (LF) at any
 redshift can be linked to the LF of local E/S0s by means of a particular
 evolutionary model. We adopted the local E/S0 LF derived by
 Kochanek et al. (2001) in the $K$-band, from the {\it 2MASS} 
survey, a Schechter function with 
$M^*_K=-24.31$, $\alpha=-0.92$ and $\phi^*=0.0015435$ $\rm Mpc^{-3}$ for
$h=0.7$. 

The ERO counts were modelled by evolving this LF with
 the galaxies split 50:50 between our elliptical and S0 models of
passive  $L^*$ evolution (see Section 4.2), 
and  including only the galaxies within
 the redshift ranges in which their respective models give ERO colours
 (i.e. approx. $z>1$).
In the Pure Luminosity Evolution (PLE) model only $L^*$ is evolved,
 while $\phi^*$ and $\alpha$ remain constant from the
 formation redshift onwards. The post-starburst fading of dsfEROs is
 assumed to be similar in degree to the passive
 evolution of the pEROs.  

In Paper I, the ELAIS:N2 ERO counts were
significantly lower than expected if all present-day E/S0s evolved by
 this PLE model, and much closer to a
non-evolving model. However, the compact angular
 sizes of most EROs
implied significant evolution in their surface brightness.

 A  merging model was also considered, which combined the PLE model's evolution
 of luminosity (per unit mass) with
 an increase with redshift in number density ($\phi^*$) and a 
corresponding decrease in
mass (parameterized as $R_{\phi}=R_{m}=0.3$; see Paper I) of the galaxies. 
The merger rate and its evolution were based on the
 observational estimate of Patton et al. (2002). This model also
 overpredicted the ERO counts. 
However, by keeping the merger model's mass evolution at
$R_{m}=0.3$ and varying the density evolution, a reasonable
best-fit to the ERO counts on ELAIS:N2 was obtained for 
 $R_{\phi}=-0.46(\pm 0.10)$. 

 In this model, described as 
`merging and negative density 
evolution' (M-DE), the comoving number density of 
passive galaxies (or any galaxies of similar colour) 
gradually decreases with redshift. Physically, this could result from
some fraction of the present-day 
massive ellipticals forming at high (e.g. $z>3$) 
redshifts, with the remainder forming
  from bluer galaxies (e.g. through interactions and 
mergers) at all intermediate redshifts.

With the deeper data available for the CDFS, we should now be able to 
improve the constraints on ERO evolution. We recompute the models of
Paper I for a $h=0.7$ Universe and ERO selection as
$I_{775}-K_s>3.92$. The M-DE model best-fitting the CDFS counts has a
density evolution parameter of $R_{\phi}=-0.49(\pm 0.06)$ (almost
identical to the best-fit model of Paper I). 

Figure 5 compares the new ERO counts with PLE,
merging ($R_{\phi}=R_{m}=0.3$), non-evolving and 
M-DE ($R_{\phi}=-0.49$) models.  
  As in Paper I, the PLE and  merging models overpredict the
observations.
 The deep CDFS counts are consistent with the
non-evolving model at brighter magnitudes, but fall below it at
$K_s>20$. This suggests that there is a real decrease in the
comoving $\phi^*$ of ERO galaxies with redshift, and not 
only the absence
of strong luminosity evolution. 

Of these models, M-DE is the only one consistent with the
plotted counts.
 In the M-DE  model the evolution of the characteristic mass (from
merging) is (as in Paper I), $m^*(z=1,2,3)=
(0.714, 0.558, 0.493)m^*(z=0)$, and the comoving number density of 
passive/red galaxies, with $R_{\phi}=-0.49$, evolves as 
 $\phi^*(z=1,2,3)=(0.577, 0.386, 0.315)m^*(z=0)$.

\section{Clustering of the EROs}
\subsection{Calculating $\omega(\theta)$} 
To investigate the clustering of EROs on the CDFS, and compare it with
other galaxies,  we calculate the
angular correlation function, $\omega(\theta)$, for (i) all
galaxies (of any colour), (ii) all
EROs, and (iii) pEROs and dsfEROs considered separately. The method adopted is
exactly as described in Paper I, with the use of $N_r=50000$ random points.

In addition, (iv) we calculate the
angular cross-correlation of the pEROs and dsfEROs, as
$$\omega_{pd}(\theta_i)={N_{pd}(\theta_i)\over N_{pr}(\theta_i)}
{N_{d}\over N_{r}}-1$$
where $N_{pd}(\theta_i)$ is the total number of pERO--dsfERO pairs
(centering on the pEROs) in the
$\theta_i$ interval of angular separation, $N_{pd}(\theta_i)$ the
number of  pERO--random point pairs, $N_{d}$ the number of dsfEROs and
$N_{r}$ the number of random points.
 As an additional check, we also evaluated the
cross-correlations with  centering on the
dsfEROs, i.e. as
$$\omega_{dp}(\theta_i)={N_{dp}(\theta_i)\over N_{dr}(\theta_i)}
{N_{p}\over N_{r}}-1$$
which gave  essentially the same results --
 $\omega_{dp}\simeq  \omega_{pd}$ at all $\theta$,  to within a
small fraction of the statistical errors, $\sim 0.1\sigma$.

The 3-dimensional two-point correlation function of galaxies
is an approximate power-law in physical separation ($r$), 
 $\xi(r)=({r\over r_0})^{-1.8}$ (Peebles 1980), at separations of
up to a few Mpc.
 Following from this, 
the $\omega(\theta)$ of galaxies will also be  an approximate power-law 
$\omega(\theta)\simeq A_{\omega}\theta^{-0.8}$, 
at the relatively small $\theta$ considered here ($\leq 0.1$ deg).
We therefore (as in Paper I) express the observed 
$\omega(\theta)$ as
 an amplitude $A_{\omega}$ at $\theta=1$ deg, obtained by
fitting with the function 
 `$A_\omega(\theta^{-0.8}-12.24)$' (where 12.24 is the
`integral constraint' as determined for the ISAAC field area; see
Paper I), over the range of separations $2<\theta<32$ arcsec. 
Error bars were estimated by the method described in Paper I, using 12
`sub-areas', and are generally a little larger than $\surd N$ errors.

\subsection{$\omega(\theta)$ Results}
Table 2  gives our $A_{\omega}$ amplitudes for a range of
magnitude limits (these results may be       less reliable at $K_s>21.5$
 due to incompleteness), 
and Figure 6 shows $\omega(\theta)$ with fitted power-laws.
We obtain statistically significant, $\sim 3\sigma$ 
 detections of  clustering, both
for all galaxies and for the EROs to $K_s=21$--22.
\begin{figure}
\psfig{file=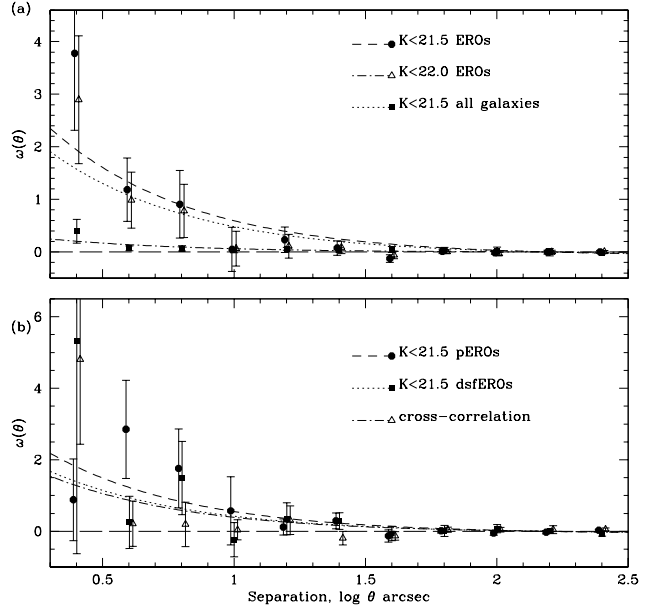,width=90mm} 
\caption{(a) Angular correlation functions $\omega(\theta)$ for
 EROs and galaxies of all colours 
on the ISAAC CDFS 
data, at the approximate completeness limit of  $K_s=21.5$, and for EROs
at the fainter $K_s=22.0$ limit, 
 with the dashed, dotted and dot-dash lines showing 
 respective best-fitting functions of the form 
 `$A_\omega(\theta^{-0.8}-12.24)$'; (b) $\omega(\theta)$ of 
 $K\leq 21.5$  EROs of pERO and dsfERO 
colour-classification, and their cross-correlation function, again
with  the dashed, dotted and dot-dash lines showing 
 best-fitting  `$A_{\omega}(\theta^{-0.8}-12.24)$'.}
\end{figure}

 Table 2 also 
gives the $A_{\omega}$ for the pERO and dsfERO subsamples.
Clustering is detected for pEROs and
consistent with a similar amplitude to
the full ERO sample, but is at best $2.3\sigma$. The detection of 
dsfERO clustering and the
cross-correlation are even weaker, and 
 would be consistent 
with this same amplitude or one only half as great.
 Clearly, more data are needed before we can make a meaningful
comparison between the clustering properties of  pEROs and  dsfEROs. 

\subsection{Interpretation of the $\omega(\theta)$}   
The observed $\omega(\theta)$ of any 
sample of galaxies will depend on their intrinsic clustering in 3D space,
described by the two-point correlation function $\xi(r)$, and their
redshift distribution $N(z)$. If $\xi(r)$ is represented by the simple
model 
  $$\xi(r,z)=(r/r_0)^{-\gamma}(1+z)^{-(3+\epsilon)}$$ 
where $r_0$ is the local correlation radius, 
$\gamma\simeq 1.8$ (observationally) and
$\epsilon$ represents the clustering evolution ($\epsilon=0$ is stable and 
$\epsilon=-1.2$ is comoving clustering), then Limber's formula (see
eg. Efstathiou et al. 1991) gives
$\omega(\theta)=A_{\omega}\theta^{-(\gamma-1)}$, where  
$$A_{\omega}=C_{\gamma} r_0^{\gamma} \int_0^{\infty}
{ (1+z)^{\gamma-(3+\epsilon)}\over
x^{\gamma-1}(z){dx(z)\over dz}}[(N(z)^2] dz/[\int_0^{\infty}N(z)dz]^2$$
where $x(z)$ is the proper distance 
and $C_{\gamma}=3.679$ for $\gamma=1.8$. 

\begin{figure}
\psfig{file=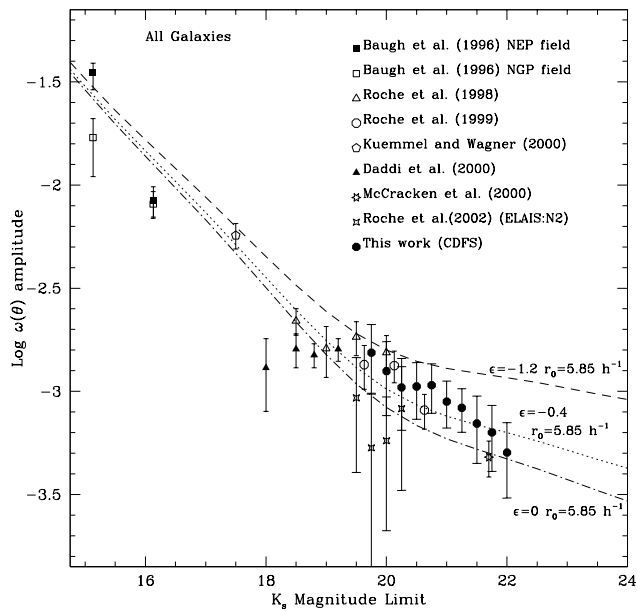,width=90mm}
\caption{The scaling of $\omega(\theta)$ amplitude with $K$ magnitude
limit, for full $K$-limited samples of galaxies from our
ISAAC CDFS data, with previously
published results, and models (all with 
$r_0=5.85 ~h_{100}^{-1}$ Mpc)
 of stable ($\epsilon=0$) and  comoving ($\epsilon=-1.2$)
clustering  and an intermediate ($\epsilon=-0.4$) evolution.}
\end{figure}
\begin{figure}
\psfig{file=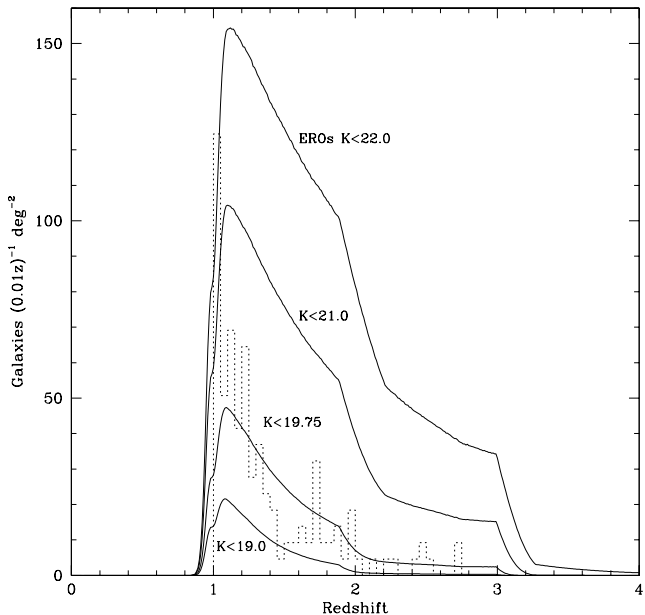,width=90mm}
\caption{$N(z)$ for $I_{775}-K_s>3.92$ EROs as given by our $R_{\phi}=-0.49$
`M-DE' model, at limits $K_s=19.0$, 19.75, 21.0 and 22.0. Also shown
for comparison (dotted histogram) is the $z>1$ tail of the Cimatti et
al. (2002b) $N(z)$ for all types of galaxy to an effective limit
$K\simeq 19.75$, normalized to the ERO model at this magnitude.}
\end{figure}
Before considering the EROs we briefly investigate
 the $\omega(\theta)$ scaling of
full $K$-limited galaxy samples. Figure 7 shows our observations for
 the CDFS, and  previously published results, which agree well.
 No correction has been applied to
 the CDFS amplitudes  for star contamination,
 but this will be very small, e.g. 
 $\Delta({\rm log}~A_{\omega})=0.04$ for the
 estimated number of stars (Section 3.4).
 Also plotted are  models based on the
 $n(z)$ from our
$R_{\phi}=R_{m}=0.3$ merging model (see Paper I), a local correlation 
radius of $r_0=5.85~h^{-1}$ Mpc (from the $I<22$  survey 
of Cabanac, de Lapparent and Hickson 2000), and showing the 
$-1.2\leq \epsilon\leq 0$ range of clustering evolution. 
The observations appear  most consistent
 with $\epsilon\simeq -0.4$--0,
 with comoving  clustering excluded.

Figure 9 shows the $A_{\omega}$ scaling of EROs selected as 
$I_{775}-K_s>3.92$ or $R-K>5.0$. Our results for the CDFS are
consistent with the scaling trend seen in previous surveys. 

Any estimation of  the 
correlation radius ($r_0$) from
$\omega(\theta)$ will inevitably be dependent on the assumed model for
$N(z)$. We represent the ERO $N(z)$ 
using the `M-DE' model (Figure 8)
Currently there is insufficient spectroscopic data for EROs to fully
test the accuracy of this $N(z)$. We can, howver, compare our model
with the
Cimatti et al. (2002b) redshift  (mostly spectroscopic, some
photometric)
 survey for all types of galaxy
at an effective limit $\simeq 19.75$. Figure 8 shows the $z>1$ tail
(138 galaxies) from this $N(z)$, which is not greatly different from our
 model and the
two may be  consistent if the fraction of EROs in
the survey  sample is highest at
$z>1.2$. 
Our model $N(z)$ is also in good agreement with the physically-based Granato et
al. (2003)  prediction for spheroidal galaxies only.

Hence, assuming our model to be a good representation of the ERO
$N(z)$, we vary the $r_0$ normalization to
best-fit (minimize $\chi^2$) 
 all the plotted data points. We estimate  $r_0$ for EROs to be $12.5(\pm
1.2)~h^{-1}$ Mpc (${\chi^2 \over N}=1.04$) for comoving and
$r_0=21.4(\pm 2.0)~h^{-1}$ Mpc (${\chi^2 \over N}=0.77$) for stable 
clustering.

\section{ERO Clusters and Pairs}
\subsection{A possible cluster of EROs}
In Paper I we noted a possible cluster of EROs in the ELAIS:N2 field, centered
on a bright ERO which is also a {\it Chandra} source and radio galaxy.
 
In the CDFS, we find another overdensity of EROs, centered on
 {\it Chandra} source XID:58 (R.A. $3^h 32^m 11.85^s$ Dec -27:46:29.14), 
 a somewhat fainter ERO ($K_s=20.94$, $H=21.66$, $J=22.92$  and 
$I_{775}=24.97$). Within a 20 arcsec radius of XID:58
there are a total of 10 $K\leq 22.0$ EROs compared with 1.37 expected
from their mean surface density on the field.
There is no accompanying overdensity of bluer galaxies, with 9 seen
 compared to 9.83 expected.

\vfill
\eject
\onecolumn
\begin{table}   
\caption{Galaxy $\omega(\theta)$ amplitudes $A_{\omega}$ (in units of
$10^{-4}$ at one degree) for full $K_s$-limited samples of
galaxies, EROs
 ($I_{775}-K_s>3.92)$, EROs classed as pEROs and dsfEROs,  and the
pERO/dsfERO cross-correlation, to a series of
magnitude limits.  $N_g$ is the number of
galaxies in each sample. The $A_{\omega}$ of EROs is  not given for the first
two limits as the samples are too small for any detection of clustering.}
\begin{tabular}{lccccccccc}
\hline
$K_s$ & \multispan{2}\hfil All galaxies \hfil & \multispan{2}
 \hfil All EROs &\multispan{2}\hfil pEROs \hfil & \multispan{2}\hfil
dsfEROs \hfil & $\rm pERO\times dsfERO$\\ \\
\smallskip
limit & $N_g$  & $A_\omega$ & $N_g$ & $A_\omega$ & $N_g$  & $A_\omega$
& $N_g$  & $A_\omega$  & $A_\omega$\\
20.00 & 467 & $12.53\pm 4.89$ & 69 &  & 41 & 28 & \\
20.50 & 638 & $10.56\pm 3.23$ & 97 &  & 56 & 41 & \\  
21.00 & 898 & $~8.92\pm 2.28$ & 137 & $53.5\pm 19.7$ & 79 & $79.4\pm
48.7$ & 58 & $72.9\pm 34.9$ & $31.3\pm 30.0$ \\
21.50 & 1233 & $~6.98\pm 2.50$ & 179 & $60.3\pm 19.0$ & 101 & $56.0\pm
24.2$ & 78 & $43.0\pm 31.0$ & $39.6\pm 21.2$ \\
22.00 & 1619 & $~5.05\pm 2.01$ & 198 & $48.9\pm 15.8$ & 109 & $45.0\pm
19.6$ & 89 & $23.5\pm 22.5$ & $31.2\pm 23.4$ \\

\hline
\end{tabular}
\end{table}

\begin{table}
\caption{EROs detected as X-ray sources, with X-ray properties from
Giacconi et al. (200) (: {\it Chandra} source XID
number, RA and Dec of the corresponding ERO 
on the $K_s$ image,
$K_s$ magnitude (total), $I_{775}-K_s$ colour (aperture),
 pERO/dsfERO colour classification (brackets denote weak, $<1\sigma$ 
classifications), 
 Soft and Hard X-ray fluxes in
units of $10^{-16}$ erg $\rm s^{-1} cm^{-2}$ (or $1\sigma$ upper
limits), and X-ray hardness ratio $HR$ (see text).}
\begin{tabular}{lcccccccc}
\hline
\smallskip
XID  & R.A & Dec. & $K_s$ & $I_{775}-K_s$ & p/dsf & $F(0.5$--$\rm
 2.0~keV)$ & $F(2$--$\rm 10~keV)$ & $HR$ \\
26 & 3:32:39.75 & -27:46:11.25 & $19.64\pm 0.09$ & $4.05\pm
0.11$  & (p) & $7.35\pm 0.750$ & $28.9\pm 4.07$ & $-0.23\pm 0.08$ \\
58 & 3:32:11.76 & -27:46:28.14 & $20.94\pm 0.17$ &  $4.04\pm
0.20$ & (dsf) & $6.36\pm 0.668$ & $18.8\pm 3.26$ & $-0.36\pm 0.09$ \\
79 & 3:32:38.04 & -27:46:26.26 &  $20.92\pm 0.11$ & $4.83\pm
0.22$ & (p) & $8.06\pm 0.785$ & $20.7\pm 3.64$ & $-0.42\pm 0.08$ \\
86 & 3:32:33.84 & -27:45:20.45 & $21.22\pm 0.16$ & $4.82\pm
0.34$ & (p) & $1.22\pm 0.345$ & $6.96\pm 2.43$ & $-0.05\pm 0.22$ \\
108 & 3:32:05.76 & -27:44:46.91 & $21.28\pm 0.12$ & $4.16\pm 0.20$ &
p & $2.94\pm 0.513 $ & $8.73\pm 2.92$ & $-0.36\pm 0.17$ \\
153 & 3:32:18.32 & -27:50:55.25 & $18.58\pm 0.04$ & $4.34\pm
0.05$ & p & $1.88\pm 0.403$ & $46.6\pm 4.43$ & $0.59\pm 0.07$ \\
188 & 3:32:22.53 & -24:49:49.75 & $18.17\pm 0.04$ & $3.94\pm
0.04$ & dsf & $<0.616$ & $6.43\pm 2.37$ & $1.00_{-0.81}$ \\
221 &  3:32:08.91 & -27:44:24.81 & $21.76\pm 0.16$ & $5.63\pm 0.84$ & (dsf) & $1.09\pm 0.360$ & $<5.12$ & $-1.00^{+0.84}$ \\
253 &   3:32:20.05 & -27:44:47.20 & $20.00\pm 0.09$ & $3.97\pm
0.11$ & (dsf) & $0.829\pm 0.320~$ & $26.0\pm 3.52$ & $0.66\pm 0.11$ \\
254 & 3:32:19.80 & -27:45:18.72 & $21.48\pm 0.21$ & $4.06\pm
0.28$ & (dsf) & $<0.573$ & $18.4\pm 3.06$ & $1.00_{-0.32}$ \\
513 & 3:32:34.01 &  -27:48:59.74 & $21.19\pm 0.17$ & $4.49\pm 0.25$
& dsf & $0.847\pm 0.331~$ & $8.60\pm 2.60$ & $0.23\pm 0.23$ \\
515 & 3:32:32.17 & -27:46:51.49 & $21.99\pm 0.19$ & $4.10\pm
0.33$ & (dsf) & $0.843\pm 0.304$ & $12.7\pm 2.74$ & $0.41\pm 0.17$ \\
572 & 3:32:22.13 & -27:48:11.47 & $21.17\pm 0.16$ & $4.92\pm
0.29$ & (dsf) & $1.00\pm 0.368$ & $<5.39$ & $-1.00^{+0.90}$ \\ 
574 & 3:32:31.53 &  -27:48:53.81 & $20.25\pm 0.11$ & $4.28\pm
0.13$ & (dsf) &  $0.787\pm 0.319$ & $<4.44$ & $-1.00^{+0.93}$ \\
600 & 3:32:13.81 &  -27:45:25.63 &  $18.66\pm 0.03$ & $4.20\pm
0.04$ & p & $<0.548$ & $18.8\pm 3.26$ & $1.00_{-0.66}$ \\
609 & 3:32:36.17 & -27:50:36.93 &  $19.64\pm 0.07$ & $4.49\pm
0.09$ & (p) & $<0.687$ & $9.31\pm 2.63$ & $1.00_{-0.65}$ \\
623 & 3:32:28.51 & -27:46:58.17 & $20.60\pm 0.09$ & $4.51\pm
0.17$ & dsf & $0.809\pm 0.300$ & $<4.49$ & $-1.00^{+0.92}$ \\    
\hline
\end{tabular}
\end{table}
\twocolumn

\begin{figure}
\psfig{file=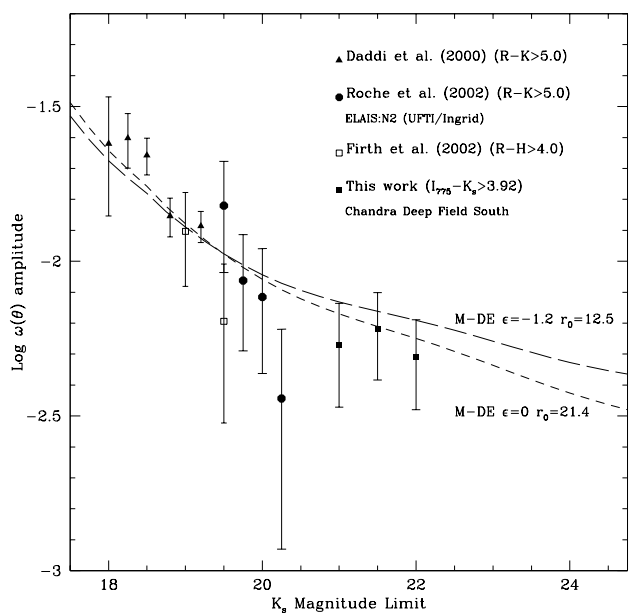,width=90mm}
\caption{The scaling of $\omega(\theta)$ amplitude with $K$ magnitude
limit for EROs selected as $I_{775}I-K_s>3.92$ or $R-K>5.0$, from our
ISAAC CDFS data and from previously
published results. These are compared with M-DE models with either stable
($\epsilon=0$)
or comoving ($\epsilon=-1.2$) clustering and $r_0$ 
best-fitted to the plotted data points.} 
\end{figure}
\begin{figure}
\psfig{file=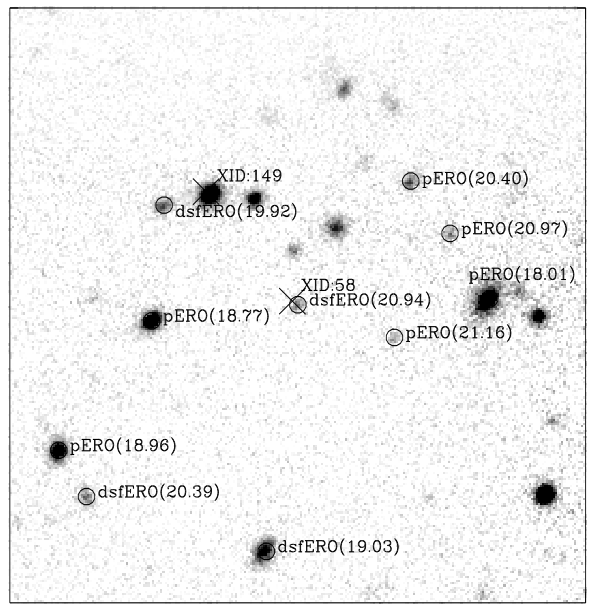,width=90mm}
\caption{Greyscale plot of the $40\times40$ arcmin region of the ISAAC
$K_s$ image centred on the {\it Chandra} source XID:58, on which X-ray source
centroids are labelled (diagonal crosses) 
with XID numbers from Giacconi et al. (2002)
and EROs (circles) by pERO/dsfERO classification and $K_s$ magnitude
(in brackets).}
\end{figure}
\begin{figure}
\psfig{file=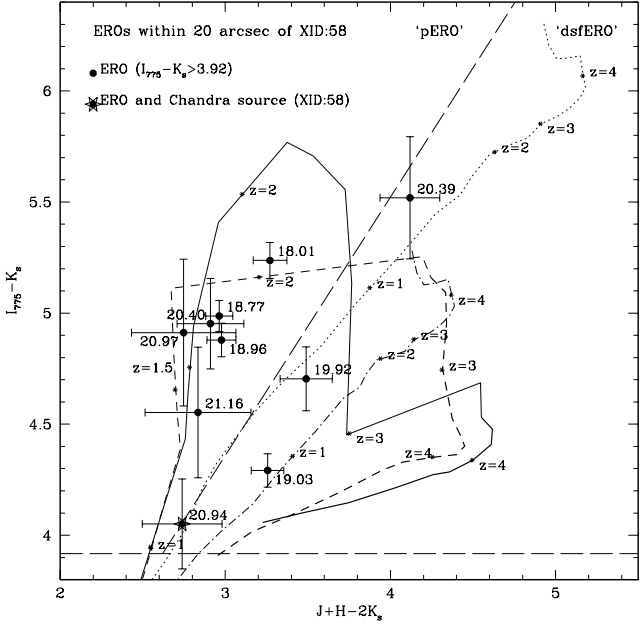,width=90mm}
\caption{Colour-colour plot of the 10 EROs within a 20 arcsec radius
of XID:58, each labelled by $K_s$ magnitude (as on Figure
10).  The long-dashed diagonal line shows the adopted pERO/dsfERO
divide, and the horizontal long-dashed line the colour limit for ERO
selection. Redshift loci for the E and S0 models are plotted as solid
and short-dash lines,  with the bolder lines indicating their high-$z$
starburst phases. Redshift loci for the dusty starburst models are
plotted as dotted (for $E(B-V)=0.8$) and dot-dash (for $E(B-V)=1.0$)
lines. Redshifts are marked on each model.}
\end{figure}
Figure 10 shows this region on the ISAAC image with
EROs and {\it Chandra} sources marked, and Figure 11 shows these 10 EROs on a
colour-colour plot.

A recent photometric redshift estimate for the XID:58
 galaxy (Mainieri et al., PhD thesis, in prep) gives $z\simeq 1.44$.
 If the overdensity is at this redshift, a 
20 arcsec radius is 0.233 Mpc for
$h=0.7$. On the colour-colour plot, six of the EROs within
 this radius 
appear clustered
 around a passive model at $z\sim 1.5$. It seems very likely that
these pEROs and the X-ray source form a single cluster. 

Of the 3
dsfEROs, one, the $K_s=20.39$ galaxy, has colours at least consistent with 
membership of this cluster. The colours of the  other two suggest lower
$z$, foreground galaxies. A
second X-ray source within the cluster area, XID:149, 
 has a spectroscopic redshift
of only 1.033 (Szokoly et al. 2003).

On Figure 11 the XID:58 galaxy is classed as a dsfERO but effectively
lies on ($0.01\sigma$ from) the
 adopted pERO/dsfERO divide. A dsfERO classification seems more
likely on the basis of (i) a $\sim 0.5$ mag 
excess in the  $I_{775}$-band ($\lambda_{rest}\simeq 320$nm)  
relative to the E/S0 model colours  at its `photo-$z$', $z=1.44$, and
(ii) the apparent morphology on the ACS image 
which does not resemble a giant elliptical, but rather is
 asymmetric and very compact.
If $z=1.44$ is assumed to be its correct redshift, we 
 estimate the $R$-band absolute magnitude as  $M_{R}\simeq -22.2$
 (for $h=0.7$), from the observed 
magnitude  closest to restframe $R$, $H=21.66$, and our passive model.
This would, in our M-DE model, be consistent
 with an $L\sim L^*$ galaxy at this redshift.

 The X-ray flux, $F(0.5$--$\rm
2.0~keV)=6.36(\pm 0.668)\times10^{-16}$ and  $F(2.0$--$\rm
10.0~keV)=1.88(\pm 0.326)\times10^{-15}$ ergs $\rm cm^{-2}s^{-1}$
would, for a $f_{\nu}\propto\nu^{-1}$ SED, corresponds to  
$L_{X}(0.5$--$\rm 10.0~keV)\simeq 10^{43.5}$ ergs $\rm s^{-1}$ for
$h=0.7$, which
is too high for a non-AGN starburst or elliptical (see Section 8).
The X-ray emission cannot be from hot intracluster gas
as it is consistent with a point-source in the 
{\it Chandra} data (Giacconi et al. 2002). The non-detection of a
diffuse component  might not be unexpected at
$z\sim 1.5$, as the X-ray luminosity
 function of clusters is known to undergo  
rapid  negative evolution  with increasing redshift (e.g. Henry 2003).
The high X-ray and modest optical luminosity of the central ERO,
 combined with the
morphology and colours all suggest it is a dusty starburst or
post-starburst galaxy (likely a recent post-merger), and the host of an
obscured AGN -- the source of the X-ray
emission.

In summary, we find evidence for the existence of a
 galaxy cluster 
 at $z\sim 1.5$, containing 7 or 8 EROs with $K\leq 22$, including an X-ray
 luminous obscured AGN (Chandra source XID:58) and several passive galaxies. 
A notable difference from the possible $z\sim 1.1$ cluster reported in
Paper I is that the X-ray source does not lie in the central giant 
 elliptical but a in a relatively small, irregular and probably dsfERO
type galaxy. 
Some of the cluster ellipticals are up to 2 mag brighter than XID:58 
and would be
 within the reach of spectroscopy, allowing the existence of a
 cluster to be verified.

\subsection{Close pairs of Galaxies, and the Merger rate}
The number of pairs of galaxies at
 very close separations, $\leq 5~{\rm  arcsec}(\sim  40~h^{-1}$ kpc
at $1<z<3$) is related primarily to the frequency
 of interactions and mergers.
Patton et al. (1992) estimated that for an optically selected sample
 of galaxies at $z\simeq 0.3$, there were $0.0321\pm 0.007$
interacting companions per galaxy, and that  this fraction and the
 merger rate were evolving
(at $z<1$) as $\propto(1+z)^{2.3}$. This result is  
 incorporated in our merging and M-DE model as a merger rate
 $R_{m}(z)=0.3(1+z)^{2.3}$ per
Hubble time. It also provides a useful point of comparison for the EROs.

We investigate the close-pair counts within the ERO sample, using
 a method originally described by Woods, Fahlman and
Richer (1995).
For each ERO-ERO pair, the probability $P$ of occurring by chance
 (in a random distribution) is estimated as 
\begin{equation}
$$P=1-{\rm exp}[-\rho(<m_2)\Omega_{\beta\theta}]$$
\end{equation}
where $\rho(<m_2)$ is the surface density of galaxies in the sample brighter
than $m_2$, the apparent magnitude 
of the fainter galaxy of the pair,

 $\Omega_{\beta\theta}$ is the area of
the annulus around the brighter galaxy between $\beta$,
 an angular separation cut-off below which individual objects cannot
be resolved ($\beta\simeq 1$ arcsec here),
 and $\theta$, the pair separation ( $\Omega_{\beta\theta}= 
\pi\theta^2-\pi\beta^2$ away from any edges or holes).
We estimate $\Omega_{\beta\theta}$  numerically by counting points
 distributed randomly over the observed field area. 

Galaxy pairs with $P<0.05$ are then counted
 as close pairs with a high probability that they are interacting.
As faint companions of brighter galaxies are especially likely to be
missed near the detection threshold, this analysis is limited to the
 $K_s\leq 21.5$ EROs. Amongst these 179, we classify 20 close pairs,       
 of which 9 are pERO--pERO, 3 are
 dsfERO--dsfERO and 8  are mixed-type --
 there is no indication that dsfEROs
 are more likely than pEROs to be in interacting pairs.
 
To estimate the number of close
 pairs resulting from chance coincidence, 
the same analysis is performed on 100
 simulations with the same number of galaxies and magnitude
 distribution as the real sample, but  randomized positions. Averaging
 over all simulations, we find a random distribution has
$8.98\pm 3.36$ pairs with $P\leq 0.05$ (where the error bar is for a
 single simulation).
\begin{figure}
\psfig{file=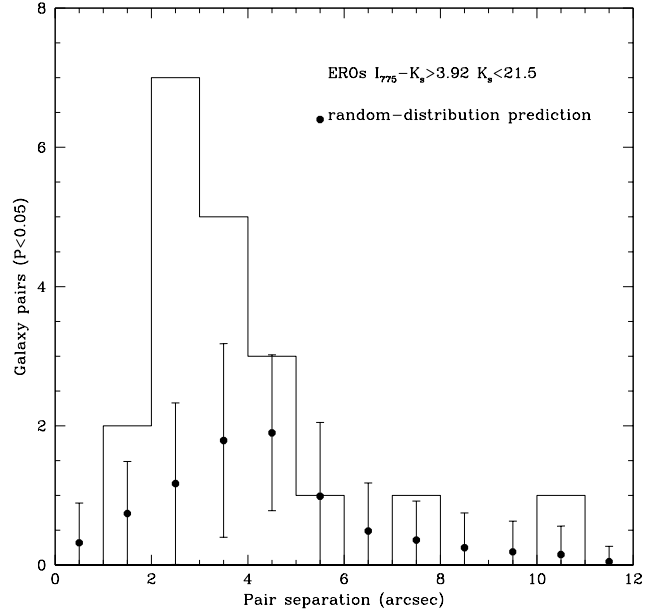,width=90mm}
\caption{Histogram of the number of close ERO-ERO pairs with $P\leq
0.05$ (as defined in the text) as a function of pair separation, as
calculated for our 179 EROs with $K_s\leq 21.5$ and $I-K_s>3.92$.  
The plotted points show the simulation prediction for the same
number of galaxies distributed randomly (where the arror bars show the
error for a single sample of 179).}
\end{figure}
 Figure 12 shows the observed and random-simulation $P\leq 0.05$ 
pair counts for $K_s\leq 21.5$ EROs 
-- the observed 
counts show an excess at 
 1--5 arcsec. 
Subtracting the random expectation from the 
 observed  pair count gives $11.02\pm 3.36$, and hence  
 the number of interacting
 ERO companions per ERO as ${11.02\pm 3.36\over 179}=0.062\pm
 0.019$. This is higher than the 
Patton et al. (2002) estimate for optically-selected
 galaxies at $z\sim 0.3$, but with $(1+z)^{2.3}$ evolution, this
 would reach 0.062 at $z=0.73$.

 However, the $P\leq 0.05$ pair count is probably
underestimated due to image merging causing some of the very closest
($\theta \leq 2$ arcsec) pairs to be missed. For a rough estimate of
this effect, we can assume the true ratio of galaxy pairs to the
random prediction at $\theta\leq 2$ arcsec is no lower than at
$2\leq \theta\leq 4 $ arcsec, where observationally it is ${12\over
3.27}$. At $\theta<2$ arcsec, we counted 2 pairs compared to the random
prediction of 1.08, but from the above ratio we should have
${12\over 3.27}\times 1.08=3.96$, an additional 1.96, increasing the 
paired fraction to  ${1.96+11.02\pm 3.36\over 179}=0.073\pm
 0.019$ companions per ERO. The Patton et al. (2002) evolution reaches
this figure at $z=0.86$, which is still less than the typical ERO
redshifts. 

On this basis we tentatively conclude that 
the incidence of visibly interacting or merging pairs
amongst EROs is comparable to,  and
 not significantly 
 greater than, that occurring in an optically-selected sample of
 galaxies  at the same $z\geq 1$ redshifts. 
\section{Correlation of EROs with X-ray Sources}  
The CDFS was surveyed with {\it Chandra} in two passbands, 0.5--2.0
keV (`Soft') and 2--10 keV (`Hard'), reaching (in a total of 942 ksec
exposure time) respective source-detection limits (for a photon index 
$\Gamma=1.375$, the mean for the detected sources) of $5.5\times
10^{-17}$ and $4.5\times 10^{-16}$ erg $\rm s^{-1} cm^{-2}$ (which are
12 and 15 counts $\rm Msec^{-1}$).
 Giacconi et al. (2002) catalog the X-ray sources, 
numbering a total of 346 detections in the Soft (307) and/or Hard (251)
bands, of which 73 are in the area of our ISAAC data.
For each source, an X-ray hardness ratio $HR$ is given as ${H-S\over H+S}$ 
where $H$ and $S$ are the photon counts in the Hard and Soft bands,
and can take any value from 
$HR=-1$ (detected only in Soft band) to $HR=+1$
(detected only in Hard). 

\subsection{Calculating the cross-correlation}
We cross-correlate these 73 X-ray sources with the EROs to
(i) identify which EROs are also X-ray
sources and (ii) investigate whether there is any
association  between EROs and X-ray sources at non-zero
separations.    

The high spatial
resolution  of {\it Chandra}, $\rm FWHM\simeq 0.7$ arcsec near the aim
point, provides
 source positions sufficiently accurate to be matched to individual
galaxies, even to $K_s=22$.
When the {\it Chandra} sources were correlated
with the $K_s$-band source list, we found $68\over 73$
  to have probable $K_s$-band IDs within $<2.5$ arcsec. We also found
a systematic offset of $\sim 1.3$ arcsec in the Giacconi et al. (2002)
 X-ray centroids. 
 To remove this, and  optimize the positional matching, we fitted
 an astrometric transform between these X-ray
 centroids (for 65 sources) and the 
  positions in pixels of
their closest $K_s$-band counterparts. 
The rms scatter in this transform was only 0.44
arcsec.

 The  {\it Chandra} source positions could then be transformed to 
accurate pixel co-ordinates on the $K_s$ image. 
The source-ERO cross-correlation
function (see Section 6.1) was evaluated as 
$$\omega_{xe}(\theta_i)={N_{xe}(\theta_i)\over N_{xr}(\theta_i)}
{N_{e}\over N_{r}}-1$$
where $N_{xe}(\theta_i)$ is the total number of {\it Chandra}--ERO pairs
(centering on the {\it Chandra} sources) in the
$\theta_i$ interval of angular separation, $N_{xr}(\theta_i)$ the
number of  {\it Chandra}--random pairs, $N_{e}$ the number of EROs and
$N_{r}$ the number of random points (50000 here).  
\subsection{X-ray detected EROs}
EROs were identified directly with {\it Chandra}
sources where the $K_s$ band and X-ray positions
coincided within 2.0 arcsec.  A total of 17 direct matches were found
(each of these was  checked by eye to verify selection of 
the  correct counterpart), of which only 1.0 would
(on the basis of the mean surface density of EROs) be expected
from chance coincidence.

If there are 17 matches, and $\sim{16\over 17}=94$ per cent are genuine, 
this means that  $8.1\pm 2.1$ per cent of the
 $K_{s}\leq 22.0$  EROs are X-ray sources above the {\it Chandra}
942 ksec detection limit. Of the 17, we classed  7 as pEROs and 10
as dsfERO, and so estimate the X-ray detected fractions of each type 
as $6.4\pm 2.4$ and 
$11.2\pm 3.6$ per cent respectively. This is higher for dsfEROs 
but the difference is not
statistically significant. These fractions are not significantly
different at $K_s\leq 21.5$ (to which there
are 14 matched EROs). Our result also means that ${17-1.0\over 68}=   
23.5\pm 5.9$ per cent of the {\it Chandra} sources detectable on the
$K_s$ image exhibit ERO colours, whether from a passive or a dusty
host galaxy.

Table 3 gives the X-ray and optical ID properties of these 17 EROs, and
Figure 13 shows total 0.5--10.0 keV fluxes
 (i.e. the sum of the 0.5--2.0 and 2.0--10.0
keV fluxes, assuming a zero flux for any band in which there is no
 detection)
against $HR$. This plot
can be compared directly with Figure 4 of Alexander et al. (2002)
and Figure 2 of Vignali et al. (2003). 

The total X-ray flux and $HR$ can help to classify
the source as an AGN, starburst or elliptical galaxy. A flat
spectrum (photon index $\Gamma=1.0$) is
$HR\simeq -0.11$. Both unobscured AGN and starburst
galaxies typically have $f_{\nu}\propto \nu^{-1}$ ($\Gamma=2.0$) 
SEDs, which is $HR\simeq
-0.35$. Obscured AGN
will have a harder $HR$, depending on the absorption column,
$N_{H}$ and redshift. Starbursts have lower X-ray luminosities 
$L_X\leq 10^{42}~h^{-2}$ ergs $\rm s^{-1}$, and hence at $z>1$, 
fluxes $F(0.5$--$10~{\rm
keV})\leq 4\times 10^{-16}$  erg $\rm s^{-1} cm^{-2}$, while AGN
tend to be more luminous.
  Passive
ellipticals, containing hot gas, 
 have a similarly soft $HR$ to the starbursts and even lower $L_X$.

\begin{figure}
\psfig{file=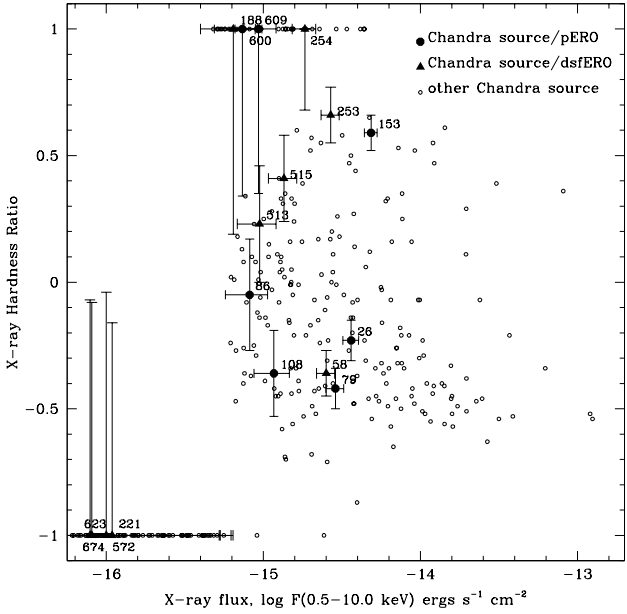,width=90mm}
\caption{Detected $H+S$ X-ray flux (0.5--10.0 keV), against hardness
 ratio $HR$), for  {\it Chandra} sources on the CDFS,
with the 17 identified with EROs indicated by large symbols 
and labelled with XID number.}
\end{figure} 

Of these 17 ERO X-ray sources, 13 (all 7 pEROs, and 6 dsfEROs) are
detected in the Hard X-ray band (9 in both bands and 4 in Hard only). These
13 have total fluxes
$F(0.5$--$\rm 10.0~keV)>6\times
10^{-16}$ ergs $\rm s^{-1}cm^{-2}$, sufficient to be  indicative of
 AGN, which must be obscured to some degree for the galaxies to 
exhibit ERO colours.
 Their $HR$s show a similar wide spread to the
X-ray detected ERO sample of Alexander et al. (2002), and -- as the
flux limits and hence (presumably) redshift ranges of these samples 
are similar --  correspond to the
same range of absorbing columns, 
$N_H\simeq 10^{21.5-23.5}\rm
cm^{-2}$.  The pERO/dsfERO mixture suggests
considerable diversity in the star formation histories of the AGN hosts.

Alexander et al. (2003) have recently found 
that many ($>36$ per cent of) bright $F_{850 \mu {\rm m}}>5$ mJy
sub-mm sources are themselves detectable in the deepest 
($\sim 1$ MSec) X-ray surveys. The X-ray luminosities of these
galaxies, $L_X\sim 10^{43}$ ergs $\rm s^{-1}$, indicate that at least
a majority contain
AGN, but are
  lower than QSO luminosities, and the sources also 
have high absorbing columns,
$N_H\geq 10^{23}~\rm cm^{-2}$. These properties would be consistent with
forming ellipticals, still in the pre-ejection, dusty phase.
It seems possible that the 5 dsfEROs in our sample with $HR>0$ 
include similar 
proto-galaxies (the surface density of sub-mm
sources detected with 1 MSec on {\it Chandra} corresponds to 5 on our
sample area), but we require sub-mm data to verify this.

We can, however, compare the X-ray properties of our EROs
with the results of Alexander et al. (2002), who in  
1 Msec {\it Chandra} survey of the CDFN detected 6 of a
total of  29
$HK^{\prime}<20.4$
29 EROs ($I-K>4.0$) in a 70.3 $\rm arcmin^2$ area.
 This $HK^{\prime}=20.4$ limit
is approximately $K_s=20.1$, and 5 of our X-ray detected EROs are
 brighter than this (Table 3). Hence, at an equivalent
combination of  X-ray and $K_s$ flux limits, 
  Alexander et al. (2002) and ourselves find consistent surface densities of
EROs, ${6 \over 70.3}=0.085\pm 0.035~\rm arcmin^{-2}$ and 
${5 \over 50.4}=0.099\pm 0.044~\rm arcmin^{-2}$
respectively. Alexander et al. (2002) also detected 1 of the 9 EROs in a
deeper sample covering 5.3 $\rm arcmin^2$, and this is also consistent
with our surface density at $K<22$.
Furthermore, our X-ray ERO samples are similar in the  proportions of 
starburst and AGN sources (on the basis of X-ray fluxes) and in their 
median hardness ratios of $HR\simeq 0$--0.3.

Vignali et al. (2003) repeated this cross-correlation on the same
field, with the addition of more  {\it Chandra} data bringing the
exposure time to 2 Msec.
 Some additional
EROs were detected, bringing the numbers to 
10 of the 29 $HK^{\prime}<20.4$
and 3 of the 9  $K<22$ EROs.

\subsection{The X-ray--ERO Correlation at $\theta>0$}
We find evidence, at the $\sim 3.3\sigma$ level,  that
 {\it Chandra} sources are also 
cross-correlated with EROs at {\it non-zero} separations.
The cross-correlation function, shown here  (Figure 14a) 
for $K_s\leq 21.5$ EROs, remains
 positive at $2\leq \theta \leq 20$ arcsec, where any signal would be
the result of clustering between separate  objects.
A function `$A_\omega(\theta^{-0.8}-12.24)$' fitted at
$2.0<\theta< 20$ arcsec,  gave an amplitude at one degree
 of $40.8\pm 12.3 \times
10^{-4}$, which is $68\pm 20$ per cent of
 the $\omega(\theta)$ amplitude (Table 2)
 for EROs at this limit. 
\begin{figure}
\psfig{file=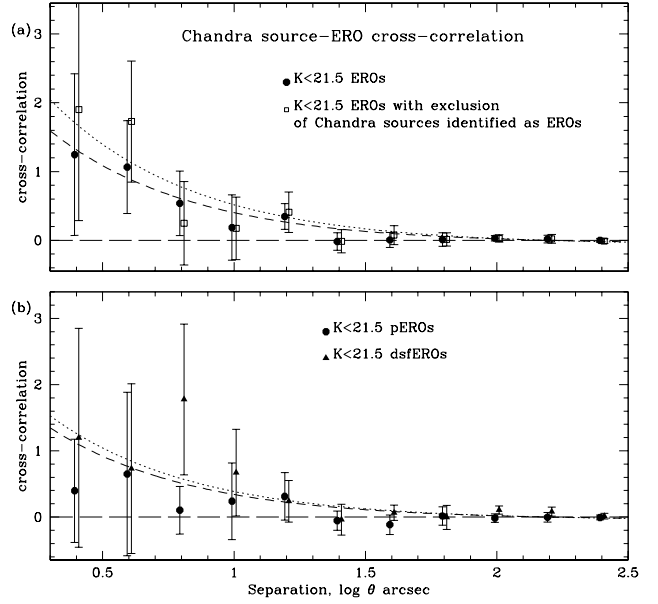,width=90mm}
\caption{(a) Cross-correlation function, at non-zero separations,
 between {\it Chandra}
sources (73) and $K\leq 21.5$  EROs (179) on the ISAAC
 imaged area of the CDFS. This is also shown
with the exclusion from the analysis 
of 17 {\it Chandra} sources identified as EROs.
 (b) Cross-correlation function
 between {\it Chandra} sources (73) and $K\leq 21.5$ EROs 
separated into pEROs (101) and dsfEROs (78).}
\end{figure}

However, it seems possible
that some of the signal at  $\theta>2$ arcsec might be due to the fact
that $\sim {1\over 4}$ of  {\it Chandra} sources are also EROs,
 and EROs are strongly 
clustered with each other.
 To investigate this,
  we excluded the 17 {\it Chandra} sources identified
as EROs and cross-correlated
only the remaining 56 with the ERO sample. The
cross-correlation falls to zero at $\theta<2$
arcsec (as expected),
 but it is little changed at $\theta>2$ arcsec, where the
best-fitting amplitude is $52.4\pm 17.4\times
10^{-4}$. This implies that, at  $z\sim 1$--2, 
  faint X-ray sources (AGN and starbursts) 
in general (i.e. not only the very reddened ones) 
cluster with EROs. 

We also calculated the cross-correlation of the  
{\it Chandra} sources with the separate pERO and
dsfERO subsamples (Figure 16b), and fitted the same function as above
at  $2.0\leq \theta\leq 20$ arcsec. At the $K_s\leq 21.5$ limit  
the best-fit $A_{\omega}$ are
 $34.5\pm 27.0\times 10^{-4}$ (pEROs) and $39.2\pm 24.2\times
10^{-4}$ (dsfEROs) --   
neither measurement  is a $2\sigma$ detection and 
there is no significant
difference between the two, so 
a larger sample will be needed to
investigate the dependence on spectral type.

Almaini et al. (2003) cross-correlated bright 
 sub-mm (SCUBA) sources and X-ray ({\it Chandra}) sources on the
 ELAIS:N2 field  and detected a signal at $4\sigma$ significance, with 
a mean cross-correlation
 function of $\sim 0.6$ at non-zero separations of $5< \theta< 100$
arcsec.
However, while the bright 
sub-mm sources are generally at high redshifts -- 
e.g. Chapman et al. (2003) find an interquartile range $1.9< z <
2.8$ for a spectroscopic sample of 10 --
 the {\it Chandra} source $N(z)$ is  peaked 
at only $z\simeq 0.7$ (Gilli et al. 2003, from
the CDFS 1 MSec survey).
 Part of the cross-correlation could be due to
gravitational lensing of the sub-mm sources by lower redshift
large-scale structure, traced by the X-ray sources (Almaini et
al. 2003b, and in prep.).

 On the other hand, the X-ray source
 $N(z)$ also has an
extended high-$z$ tail, with $\sim 20$ per cent 
at $z>1.5$. Furthermore these high-$z$ X-ray sources are clustered, with
$N(z)$ spikes at $z=1.618$ and
$z=2.572$, and some of the faintest {\it Chandra} sources are
themselves {\it SCUBA} detections (Alexander et al. 2003).
 Hence, 
it is probable that a real clustering together of sub-mm and X-ray
sources is also contributing. 

Fortunately, lensing effects  will be much less
significant in the case of 
our {\it Chandra} source -- ERO correlation. Firstly, the effect 
of lensing on the observed surface density of any type of source
depends on its
 number count slope; for a power-law ${dN\over
 dm}=\gamma$ the cumulative count to magnitude $m$ 
is modified by a muliplicative factor $\mu^{2.5\gamma - 1}$, where $\mu$
is the lensing magnification (Almaini et al. 2003b).
 The sub-mm number counts are extremely
steep, $\gamma\simeq 1$, whereas the majority of the EROs are at
$K_s>19.25$ with  $\gamma\simeq 0.16$. Lensing of
$\mu\simeq 1.25$ near a cluster would then enhance the probability of
finding a sub-mm source by 40 per cent but {\it reduce} the number of EROs
by 10 per cent. 
Secondly, the redshifts of  EROs 
 (e.g. Cimatti et al. 2002a, 2002b) show much more overlap with the
{\it Chandra} sources than do the sub-mm sources.
Hence, we  interpret our cross-correlation as evidence that these 
galaxies do trace the same large-scale structures.

\section{Summary and Discussion}
\noindent {\bf 1.} We identify a sample of 198 EROs, 
defined here as $I_{775}-K_s>3.92$ galaxies, to a limit $K_s=22$ 
 on public ESO/GOODS data covering 50.4
   $\rm arcmin^2$ of the Chandra Deep Field South (CDFS). Of these, 179
are brighter than an estimated completeness limit of $K_s=21.5$.

The number counts of EROs flatten
markedly at $K_s\simeq 19.0$--19.5, from  
$\gamma\simeq 0.59\pm 0.11$ to $\gamma\simeq 0.16\pm 0.05$.
As we previously reported, counts of  EROs are significantly lower than
 predicted by a model in which all present-day E/S0 galaxies have undergone
 pure luminosity evolution (PLE).
  At $K_s>20$, the ERO counts fall below even a
 non-evolving E/S0 model, suggesting that the 
comoving number
density (and not only the mass or luminosity) of red/passive galaxies  
is lower at $z\geq 1$.

The ERO counts can be fitted much more closely by our `merging and
negative density evolution' (M-DE) model (Paper I),
 in which a local 
luminosity function for E/S0 galaxies 
is evolved through a combination of (i) passive $L^*$
evolution, (ii) merging at an evolving rate based on an 
observational estimate (Patton et al. 2002), and (iii) a gradual decrease
with redshift in the comoving number density of red/passsive galaxies,
 parameterized as $R_{\phi}$.
The best-fitting
$R_{\phi}\simeq -0.49(\pm 0.06)$ corresponds to 
a 42/61/68 per cent reduction in $\phi^*$ at $z=1/2/3$.
 This could be interpreted as a $\sim {1\over 3}$ fraction of
the present day comoving number density of 
 E/S0 galaxies forming  at $z>3$ and the remainder
from mergers and  interactions of bluer galaxies (e.g. spirals) over  
all intermediate redshifts.

\medskip

\noindent {\bf 2.} We investigate the clustering of the CDFS EROs and
detect a $>3\sigma$ signal in the angular correlation function,
$\omega(\theta)$, at $K_s=21-22$ limits. The
$\omega(\theta)$ amplitudes, combined  with those from 
previous studies of ERO clustering
(Paper I; Daddi et al. 2000; Firth et al. 2002), are interpreted using
models based on the Limber's formula integration of our
 `M-DE' model $N(z)$ for EROs. The 
$\omega(\theta)$ scaling of EROs at $K=18$--22 limits is
 well-fitted with either comoving
($\epsilon=-1.2$) clustering and a correlation radius $r_0=12.5(\pm
1.2)~h^{-1}$ Mpc, or stable clustering ($\epsilon=0$) and 
 $r_0=21.4(\pm 2.0)~h^{-1}$ Mpc.
Our sample appears to be insufficiently large 
for us to either confirm or
 exclude the recent claim that pEROs are more clustered at $z\simeq 1$
than dsfEROs (Daddi et al. 2002).

For full $K$-limited samples of galaxies, $\omega(\theta)$ is
consistent the locally measured correlation radius 
of $r_0=5.85~h^{-1}$ Mpc, with little or no clustering evolution, 
$\epsilon\simeq -0.4$--0. In contrast, EROs are even more
clustered than local giant ($L>L^*$) 
ellipticals (with $r_0\simeq 8~h^{-1}$ Mpc), implying that if the EROs are 
 direct progenitors of the
E/S0s,  they undergo
strong clustering evolution ($\epsilon\leq -1.2$).
 
This could be part of the same process
as the $\phi^*$ evolution, if the oldest EROs form as strongly
clustered, massive sub-mm
sources, 
while the younger EROs, added at lower redshift, form 
 from less strongly clustered disk galaxies and progessively dilute the
red-galaxy $\xi(r)$ down to its present-day value. 
This  scenario may be supported by  a 
a preliminary `COMBO 17'
finding (Phleps and
Meisenheimer 2002) that  E--Sb type galaxies 
show a rapid increase of clustering with redshift
($\epsilon=-1.35\pm 0.24$)
over the $0<z<1$  range.              

The form of ERO clustering evolution should be better constrained in
the near future as  our planned spectroscopic survey of these and
other EROs will allow investigation of ERO clustering
 in three-dimensions, and  
 forthcoming clustering estimates for the sub-mm galaxies
(Percival et al. 2003) will
provide an essential high-$z$ point of comparison.

\medskip

\noindent {\bf 3.} There is an overdensity of EROs centered on the {\it
Chandra} X-ray source XID:58, a $K_s=20.94$ galaxy and itself an ERO,
at R.A. $3^h 32^m 11.85^s$, Dec -27:46:29.14. Within 20 arcsec of this
object, we find 10 $K_s<22$ EROs, compared to 1.37 expected by
chance. Of these, 7 or 8 have colours consistent with $z\sim 1.44$, a
recent photometric estimate for the {\it Chandra} source.

 The point-source X-ray emission from XID:58 corresponds to a 
luminosity  $L_{X}(0.5$--$\rm 10.0~keV)\simeq 10^{43.5}$ ergs $\rm s^{-1}$ for
$h=0.7$, which  indicates the presence of an obscured AGN.
 The host galaxy is of a
 modest optical luminosity -- we estimate $M_R\simeq -22.2$,
 which in our model would be near $L^*$ at this redshift. 
  Willott et al. (2002) found an $L\sim L^*$
pERO in ELAIS:N2 (object N2:28) with a very similar 
X-ray to $K$-band flux ratio. The host galaxy appears compact and irregular
and we consider it more likely to be a dsfERO type, probably a recent
post-merger, than a giant elliptical.

\medskip

\noindent {\bf 4.} We investigate the numbers of close (i.e. probably
interacting) pairs of EROs, using the method of Woods, Fahlman and
Richer (1995) with the same pair selection criterion 
($P\leq 0.05$). From the close pair counts we estimate the
number of interacting ERO companions per ERO (to $K_s=21.5$) as
 $0.062\pm 0.019$. This will a lower limit as it is probable that some 
 companions will  be missed, e.g.  due to merging of the images. A
very approximate
correction for this revises our estimate up to $\sim 0.073\pm 0.019$.

Patton et al. (2002) estimated that normal, optically-selected
galaxies at $z\simeq 0.3$ had $0.0321\pm
0.007$ interacting companions per galaxy, this evolving 
as $\propto(1+z)^{2.3}$ out to
$z\sim 1$. Hence, our result suggests that the pair fraction amongst
EROs is
 higher than found in
 normal, optically-selected galaxies at $z\leq 0.3$, but consistent
with these at $z\geq 1$,  even for our
higher estimate.
  In contrast, radio-selected faint galaxies
(primarily very strong starbursts) have a much higher
 interacting fraction,
$\sim 30$--60 per cent (Windhorst et
al. 1995; Roche, Lowenthal and Koo 2002a),  reflecting the importance of tidal
   triggering in producing the highest star formation rates.

If the dsfEROs are undergoing disk-to-elliptical
transformations, we might expect at least this type of ERO
 to show an elevated merger fraction.  However, we previously (Paper I) 
estimated,  from the  high dsfEROs/pERO
ratio, that the dsfERO stage has a mean lifetime of $\sim 1$ Gyr, 
much longer than the timescales of either  physical
coalescence or merger-triggered
starbursts ($\sim 0.1$ Gyr). Furthermore,
  simulations  (Bekki and Shioya 2000) 
predict the most dust-reddened mergers will be
those occurring with a retrograde-retrograde geometry of the galaxy
rotation axes, and that these
will starburst at a late stage of merging and 
lack obvious tidal tails. 

On this basis,  the majority of the mergers showing
dsfERO colours may  be already be 
in post-burst, post-coalescence   stages,and in Section 7  would
have been counted  as single
galaxies rather than as pairs. If this is the case, many of the
dsfEROs may be asymmetric or show other morphological features
indicative of recent merging. We shall investigate  the morphologies of
these EROs in   a forthcoming paper (Caputi et al., in prep).

\medskip

\noindent {\bf 5.}  We find 17 coincidences
between the 73  {\it Chandra} X-ray sources  and the 198 EROs 
($K_s\leq 22$ $I-K_s>3.75$) in our data area, and from this
 estimate that $8.1\pm 2.1$ per cent of the
 $K_{s}\leq 22.0$  EROs are X-ray sources above the {\it Chandra}
942 ksec detection limit. 
Of the 17 detected EROs, we classify 7 and pEROs and 10 as dsfEROs. 13
of the 17 are detected in the Hard band and have total X-ray fluxes 
 sufficient to require the presence of AGN. These
have a wide range of X-ray hardness ratios consistent with AGN
 obscuring columns
$N_H\simeq 10^{21.5-23.5}\rm
cm^{-2}$. The other four EROs, detected only in the
Soft band (all  dsfEROs), may be non-AGN
starburst galaxies.

The numbers and properties of the X-ray detected EROs are consistent
 with and very similar to the results of Alexander et al. (2002) for
 the CDFN. 
 The results of Vignali et al. (2003) suggest that doubling the {it
 Chandra} exposure to 2 Msec would reveal a further
$\sim 10$ per cent of the EROs to be faint X-ray sources.

\medskip

\noindent {\bf 6.} We also find a 
 cross-correlation  (at $\sim 3.3\sigma$ significance) between
{\it Chandra} sources and EROs at non-zero ($\sim 2$--20 arcsec)
separations. We estimate the cross-correlation has an 
amplitude $68\pm 20$ per cent that of ERO $\omega(\theta)$. 
Our sample is insufficient to determine the relative contributions of
pEROs and dsfEROs.

The cross-correlation 
is not significantly reduced when the 17 {\it Chandra}
sources coincident with EROs are excluded. This implies that EROs are
clustered with faint X-ray sources at $z\sim 1$--2, ie. with AGN and
 powerful starbursts,  in general and not
only with those that are heavily reddened. 
As discussed in the Introduction, this is as expected
 for the scenario (Archibald et al. 2002; Granato et
al. 2003) where sub-mm sources evolve into EROs via a
QSO phase.

We plan to use ultra-deep spectroscopy of these EROs and others to
investigate further (and in three dimensions),  the environments  and 
cross-correlations
of pEROs, dsfEROs and faint X-ray sources.

\subsection*{Acknowledgements}
This paper is based on 
observations with the ANTU Very Large Telescope, operated by the
European Southern Observatory at Cerro Paranal, Chil\'{e}, and forming
part of the  publically available ESO/GOODS dataset.
  
NR acknowledges the support of a PPARC Research Associateship. 
OA  acknowledges the support of a Royal Society  Research Fellowship.
JSD acknowledges the 
support of a PPARC Senior Fellowship.  
  
\end{document}